\documentclass[aps,prd,a4paper,nofootinbib,
preprintnumbers,superscriptaddress,showpacs,showkeys]{revtex4-2}

\usepackage{amsmath}
\usepackage{cleveref}
\usepackage{amssymb}
\usepackage{bm}
\usepackage{graphicx}
\usepackage{xcolor}
\usepackage{slashed}
\usepackage{comment}
\usepackage[normalem]{ulem}

\crefname{section}{Sec.\!}{Secs.\!}
\crefname{equation}{Eq.\!}{Eqs.\!}
\crefname{figure}{Fig.\!}{Figs.\!}
\crefname{table}{Tab.\!}{Tabs.\!}
\crefname{appendix}{App.\!}{Apps.\!}
\newcommand{\be}{\begin{equation}}
\newcommand{\ee}{\end{equation}}
\newcommand{\ba}{\begin{eqnarray}}
\newcommand{\ea}{\end{eqnarray}}

\begin{document}

\title{Topological susceptibility and axion potential
in two-flavor superconductive quark matter}

\author{Fabrizio Murgana}\email{fabrizio.murgana@dfa.unict.it}
\affiliation{Department of Physics and Astronomy "Ettore Majorana", University of Catania, Via S. Sofia 64, I-95123 Catania, Italy}\affiliation{INFN-Sezione di Catania, Via S. Sofia 64, I-95123 Catania, Italy}

\author{David E. Alvarez Castillo}\email{dalvarez@ifj.edu.pl }
\affiliation{Institute of Nuclear Physics, Polish Academy of Sciences, Radzikowskiego 152, 31-342 Cracow, Poland}
\affiliation{Incubator of Scientific Excellence - Centre for Simulations of Superdense Fluids, University of Wroclaw, plac Maksa Borna 9, PL-50204 Wroclaw, Poland}
\affiliation{Facultad de Ciencias Físico Matemáticas, Universidad Autónoma de Nuevo León, Av. Universidad S/N, C.U., 66455 San Nicolás de los Garza, N.L., Mexico}

\author{Ana G. Grunfeld}\email{ag.grunfeld@conicet.gov.ar}
\affiliation{CONICET, Godoy Cruz 2290,  C1425FQB Ciudad Aut\'onoma de Buenos Aires, Argentina}
\affiliation{Departamento de F\'\i sica, Comisi\'on Nacional de Energ\'{\i}a At\'omica, Avenida Libertador 8250,
	C1429 BNP, Ciudad Aut\'onoma de Buenos Aires, Argentina}

\author{Marco Ruggieri}\email{marco.ruggieri@dfa.unict.it}
\affiliation{Department of Physics and Astronomy "Ettore Majorana", University of Catania, Via S. Sofia 64, I-95123 Catania, Italy}\affiliation{INFN-Sezione di Catania, Via S. Sofia 64, I-95123 Catania, Italy}

%\vspace{10pt}
%\begin{indented}
%\item[]February 2016
%\end{indented}

\begin{abstract}
We study the potential of the axion, $a$, of Quantum Chromodynamics, 
in the two-flavor color superconducting phase 
of cold and dense
quark matter. We adopt a Nambu-Jona-Lasinio-like model.
Our interaction contains two terms,
one preserving and one breaking the $U(1)_A$ symmetry:
the latter is responsible of the coupling of axions to quarks.
We introduce two quark condensates, $h_L$ and $h_R$,
describing condensation for left-handed 
and right-handed quarks respectively;
we then study the loci of the minima of the thermodynamic potential,
$\Omega$, in the $(h_L,h_R)$ plane, noticing how the instanton-induced
interaction favors condensation in the scalar channel when the
$\theta-$angle, $\theta=a/f_a$, vanishes. 
Increasing $\theta$ we find a phase transition 
where the scalar condensate rotates into a pseudo-scalar one.
We present an analytical result for the topological susceptibility,
$\chi$, in the superconductive phase, which stands both at zero
and at finite temperature. Finally, we compute the axion mass and
its self-coupling. In particular, the axion mass $m_a$
is related to the full
topological susceptibility via $\chi=m_a^2 f_a^2$, hence our result for
$\chi$ gives an analytical result for $m_a$ in the superconductive
phase of high-density Quantum Chromodynamics.
\end{abstract}



\maketitle

 %\pagecolor{black}
 
\section{Introduction\label{sec:intro}}

Axions, originally proposed by Peccei and Quinn in 1977 as a possible solution to
the strong CP problem in Quantum Chromodynamics (QCD) \cite{Peccei:1977ur,Peccei:1977hh,Weinberg:1977ma,Wilczek:1977pj,Kim:2008hd,Easson:2011zy,Berkowitz:2015aua,GrillidiCortona:2015jxo,Davidson:1981zd}, have since then
become prime candidates for dark matter \cite{Weinberg:1977ma, Duffy:2009ig,Turner:1990uz,Visinelli:2009zm, {Sikivie:1982qv}, {Kim:1979if}, {Shifman:1979if}, {Dine:1981rt}, Preskill:1982cy,Abbott:1982af,Dine:1982ah}. The quest to understand the nature of dark matter, which constitutes a significant 
portion of the mass of the universe, has led to extensive theoretical and 
experimental efforts, with axions emerging as particularly compelling 
candidates due to their unique properties. Theoretical models and experiments constrain the axion mass to be very light, on the order of $10^{-6}$ to $10^{-3}$ eV~\cite{Caputo:2024oqc}. This range arises from considerations such as the possible violation of the CP symmetry of QCD and the consequent electric dipole moment for the neutron \cite{Crewther:1979pi,Baker:2006ts,Griffith:2009zz,Parker:2015yka,Graner:2016ses,Yamanaka:2016itb,Guo:2015tla,Bhattacharya:2015esa}, astrophysical and cosmological observations, and experimental searches for axions using techniques like cavity haloscopes and axion helioscopes.
In addition to their potential role as dark matter candidates, axions, behaving as scalar fields, may also serve as the constituents of boson stars \cite{Colpi:1986ye} and arrange into axion stars \cite{Tkachev:1991ka,Kolb:1993zz,Chavanis:2011zi,Guzman:2006yc,Barranco:2010ib,Braaten:2015eeu,Davidson:2016uok,Eby:2016cnq,Helfer2017black,Levkov:2016rkk,Eby:2017xrr,Visinelli:2017ooc,Chavanis:2016dab,Cotner:2016aaq,Bai:2016wpg} and form Bose-Einstein condensates \cite{Sikivie:2009qn,Chavanis:2017loo}, see also \cite{Balkin:2022qer,Balkin:2023xtr} for further astrophysical applications. 

Furthermore, not only axions but axion-like particles (ALP) have been proposed to play the role of dark matter in order to account for missing matter at cosmological scales. It has been speculated that axions may exist in compact star interiors, and can be buried deep in their cores, giving rise to interaction with quark matter, as we investigate in this study. In the assumption that axions may be able to escape compact star interiors, they would be prone to interact with magnetic fields via the so called Primakov effect, i.e., the axion resonantly converting into a radio photon~\cite{Wang:2021wae,Calore:2021hhn}, which would provide a feasible way to detect them. The emission of axions could also cool compact stars, deviating their thermal evolution from standard scenarios~\cite{Sedrakian:2015krq,Leinson:2014ioa,Sedrakian:2018kdm,Buschmann:2021juv}. 

Macroscopic properties of compact stars bearing axions have been presented in~\cite{Lopes:2022efy, Lenzi:2022ypb}, where particular modeling has resulted in unstable compact stars that would collapse due to radial oscillations if vector repulsive interactions in quark matter are not taken into account.
Furthermore, axions have been investigated within astrophysical contexts, particularly in relation to supernova explosions and the formation of protoneutron stars ~\cite{Lucente:2020whw,Fischer:2021jfm}.
Previous works on the coupling of the QCD axion
to quarks can be found in~\cite{Nambu:1961fr,Nambu:1961tp,Klevansky:1992qe,Hatsuda:1994pi,Buballa:2003qv,Ruester:2005jc,Blaschke:2005uj,Lu:2018ukl,Lopes:2022efy,Bandyopadhyay:2019pml,Das:2020pjg,Mohapatra:2022wvj},
where in particular the effect of the QCD chiral phase transition on the low-energy properties of the axion itself has been investigated,
both at finite temperature and at finite quark chemical
potential, $\mu$.
The axions enter the model similarly to the
$\theta-$angle of QCD; in fact, formally one can pass
from QCD at finite $\theta$ to QCD with a finite axion background by identifying $\theta=a/f_a$, where $a$
denotes the axion field and $f_a$ is the axion decay
constant. Therefore, studies of the QCD interaction with 
axions at finite chemical potential serve 
as studies of quark matter at finite $\theta$ and finite $\mu$
as well. Hence, within our work we 
will interchangeably use $\theta$ in place of $a/f_a$.

Deep within the dense cores of compact stars, where matter is subjected to extreme pressures and temperatures, exotic phases of strongly interacting matter come into play, as the 2SC color superconductive phase~\cite{Alford:2007xm,Shovkovy:2004me,Buballa:2003qv,Rapp:1999qa,Rapp:1997zu}. 
This phase is characterized by a non-vanishing
quark-quark condensate: as such, it is not a color singlet,
transforming as a color-antitriplet under color rotations.
In most previous calculations, in particular in those
related to the QCD phase diagram, only the scalar condensate
was considered, as it is the one that is favored by the
one-instanton exchange. Within the present work,
we introduce both a scalar and a pseudo-scalar condensate,
since both of them are relevant when the coupling to the
axions is considered. In agreement with the common lore,
we show that our model is consistent with favoring the
scalar condensate when the axion is not included in the 
model. On the other hand, changing $\theta$ can result
in a phase transition to a new ground state where
the condensate is a pseudo-scalar one, in 
qualitative agreement with previous studies in normal
quark matter~\cite{Lu:2018ukl,Lopes:2022efy,Bandyopadhyay:2019pml,Das:2020pjg,Mohapatra:2022wvj}.
Therefore, in principle both condensates need to be
introduced. 

In the low-energy regime, where non-perturbative effects dominate, effective models become indispensable tools for describing strongly interacting matter. Chiral Perturbation Theory ($\chi$PT) stands out as a frequently employed effective framework, significantly contributing to the understanding of the vacuum structure of  QCD and the properties of axions at low temperatures \cite{Brower:2003yx,Mao:2009sy,Aoki:2009mx,Bernard:2012ci,Bernard:2012fw,Metlitski:2005di}. $\chi$PT demonstrates notable advantages in low-energy scenarios; for instance, its prediction of topological susceptibility at zero temperature aligns well with lattice QCD findings \cite{Borsanyi:2016ksw,Aoki:2017imx,Bonati:2015vqz}. However, its applicability diminishes at high temperatures and/or large densities, as it lacks information about QCD phase transitions. Hence, there's a need for a QCD-like model capable of accommodating axions and capturing the QCD phase transition dynamics. One of the most popular is the Nambu-Jona-Lasinio (NJL) model, which describes the dynamics of fermions and their interactions through effective four-fermion interactions \cite{Nambu:1961tp,Nambu:1961fr}.
 {In the formulation of the model used in the present work it includes an
instanton-induced interaction,
} which breaks the $U(1)_A$ symmetry, 
and it can describe both the spontaneous breaking of chiral symmetry and how quarks interact with axions. 

The interaction of quark matter with axions has been already explored in~\cite{Zhang:2023lij,Lu:2018ukl,Das:2020pjg,Lopes:2022efy,Bandyopadhyay:2019pml,Mohapatra:2022wvj,Gong:2024cwc}, in the framework of the NJL model, where the effect of the chiral phase transition on the properties of the QCD axion was explored. Furthermore, the axion potential was studied at finite quark chemical potential, and the behavior of axion domain walls \cite{Gabadadze:2000vw,Sikivie:1982qv,Takahashi:2018tdu, ParticleDataGroup:2022pth,Davidson:1984ik,Davidson:1983tp, Gabadadze:2000vw} (see also~\cite{Nagashima:2014tva,Shifman:2022shi} for 
a pedagogical introduction to walls) 
in bulk quark matter was also investigated. 
It was found that the axion potential is very sensitive to
the chiral phase transition, particularly when quark
matter is near criticality, where the axion mass
decreases and the self-coupling is enhanced.
In this study, we aim to extend our previous work~\cite{Zhang:2023lij} considering the coupling of axions to diquarks, in order to take into account the effects of the presence of axions in a color-superconducting medium. 
Our model is based on a four-fermion effective interaction
that contains a $U(1)_A-$preserving term,
that can be interpreted as an effective way to describe
one-gluon exchange, as well as a 
$U(1)_A-$breaking term, describing the instanton-mediated
effective interaction. The relative strength of the two
terms is regulated by a dimensionless,
free parameter of the model,
$\zeta$, while the overall strength of the effective coupling,
$G_D$,
is chosen in order to reproduce a given value of the
superconductive gap when $\zeta$ is varied.
A previous study of the coupling of the axion to a color-supreconductive phase, based on chiral effective theories, can be found in \cite{Balkin:2020dsr}.

Our main results are related to the study of the 
full topological susceptibility, $\chi=\partial^2\Omega/\partial\theta^2$, in the color-superconductive phase. Here, $\Omega$ denotes the full
thermodynamic potential of superconductive quark matter,
while the derivative is understood at $\theta=0$.
We derive, within the model at hand,
an exact analytical relation between $\chi$ and the 
physical value of the superconductive gap. 
$\chi$ is related to the axion mass, $m_a$,
and decay constant, $f_a$, via
$\chi=m_a^2 f_a^2$, therefore 
the knowledge of $\chi$ allows us to extract an explicit,
analytical formula for $m_a$ in the 2SC phase.
Moreover,
we compute the full axion potential, 
and extract from it the low-energy parameters of this potential: besides $m_a$, we compute the axion self-coupling,
$\lambda$.
Within this work, we limit ourselves to a one-loop
approximation (usually called the mean field approximation),
leaving the study of the role of quantum fluctuations to
future works.

The plan of the article is as follows. In section II we 
describe the model we adopt in our work. In section III we
present our results regarding the superconductive gap 
at finite $\theta$ as well as the axion potential in the
two-flavor superconductive phase. Finally, in section IV we
draw our conclusions and discuss  possible future works. We use natural units $\hbar=c=k_B=1$ throughout this article.

\section{The model\label{sec:model}}
In this section, we present in some detail the model we use in our
work. We firstly present the Lagrangian density, describing
how we couple the axion to quarks in the superconductive phase
of QCD. Next we turn to write the thermodynamic potential, computed
at one loop. We then show the shape of this potential
in the gaps space, discussing the location of its minima: this
discussion is useful to understand the results we show in the
next section. Finally, we present an approximated solution
to the gap equation, valid for small $\theta=a/f_a$, 
that allows us to qualitatively understand the behavior of the
superconductive gap versus $\theta$.

\subsection{Lagrangian density}

To begin with, we consider the Lagrangian density
\begin{eqnarray}
\mathcal{L}_\mathrm{int} &=&
 g_1 (q^T C i\gamma_5 
 \varepsilon\varepsilon q) 
 (\bar q i \gamma_5 C\varepsilon\varepsilon\bar q^T)
 + g_2 (q^T C  \varepsilon\varepsilon q) 
 (\bar q  C   \varepsilon\varepsilon\bar  q^T).
\label{eq:startingfrom}
\end{eqnarray}
This has been used in many works on the superconductive phases
of QCD, see for example~\cite{Blaschke:2005uj,Ruester:2005jc};
in those references, it is assumed that $g_1 = g_2$ so the
$U(1)_A$ symmetry is preserved at the Lagrangian level. 
In this work, we assume $g_1 \neq g_2$ from the very beginning,
in order to break
the $U(1)_A$ symmetry at the level of the Lagrangian
mimicking the instanton-mediated interaction.
In the Lagrangian density~\eqref{eq:startingfrom}
we used $C=-i\gamma_2\gamma_0$, satisfying $C^2=C^\dagger C=-1$.
Moreover, we adopted a condensed notation that suppresses the color
and flavor indices carried by the fields and the antisymmetric
symbols in every bilinear. For example,
\begin{equation}
q^T C i\gamma_5 
 \varepsilon\varepsilon q
 =
q^T_{\alpha i} C i\gamma_5 
 \varepsilon_{\alpha\beta 3}\varepsilon_{ij3} q_{\beta j};
\end{equation}
here $\alpha$, $\beta$ denote color indices, while $i$, $j$ stand for
flavor indices.

Starting from~\eqref{eq:startingfrom}, 
we isolate 
a term that is invariant under $U(1)_A$ and a term that explicitly
breaks this symmetry. This can be easily achieved
by
adding and subtracting the terms $ g_2 (q^T C i\gamma_5 
 \varepsilon\varepsilon q) 
 (\bar q i \gamma_5 C\varepsilon\varepsilon\bar q^T)$ and $ g_1 (q^T C  \varepsilon\varepsilon q) 
 (\bar q  C   \varepsilon\varepsilon\bar  q^T)$; we then get
 \begin{eqnarray}
\mathcal{L}_\mathrm{int} &=&
 \frac{(g_1+g_2)}{2} \left[(q^T C i\gamma_5 
 \varepsilon\varepsilon q) 
 (\bar q i \gamma_5 C\varepsilon\varepsilon\bar q^T)
 + (q^T C  \varepsilon\varepsilon q) 
 (\bar q  C   \varepsilon\varepsilon\bar  q^T)\right] \nonumber\\
 &+&\frac{(g_1-g_2)}{2} \left[(q^T C i\gamma_5 
 \varepsilon\varepsilon q) 
 (\bar q i \gamma_5 C\varepsilon\varepsilon\bar q^T)
 - (q^T C  \varepsilon\varepsilon q) 
 (\bar q  C   \varepsilon\varepsilon\bar  q^T)\right].
 \label{eq:intermediate_1}
\end{eqnarray}
The first line in the above Lagrangian density is now $U(1)_A-$preserving, so we confined the 
breaking of $U(1)_A$ to the second line of~\eqref{eq:intermediate_1}.
Writing $q=(\mathcal{P}_L+\mathcal{P}_R)q$, where
\begin{equation}
\mathcal{P}_R = \frac{1+\gamma_5}{2},~~~
\mathcal{P}_L = \frac{1-\gamma_5}{2},
\label{eq:projectors}
\end{equation}
we can easily rewrite the second line of~\eqref{eq:intermediate_1} as
 \begin{equation}
(g_1-g_2) \left[
(q^T C i\gamma_5 \mathcal{P}_L \varepsilon\varepsilon q)
(\bar q   i\gamma_5  \mathcal{P}_L C \varepsilon\varepsilon
\bar  q^T) + 
(q^T C i\gamma_5 \mathcal{P}_R \varepsilon\varepsilon
q) 
(\bar q i \gamma_5 \mathcal{P}_R C \varepsilon\varepsilon
\bar q^T)\right],
 \end{equation}
where we used $\gamma_5\mathcal{P}_L q=-\mathcal{P}_L q=-q_L$ and
$\gamma_5\mathcal{P}_R q=\mathcal{P}_R q=q_R$. Finally, defining 
$G_D=(g_1 + g_2)/2$ and $\zeta G_D=(g_1 - g_2)$, we can rewrite the
interaction~\eqref{eq:startingfrom} as
\begin{eqnarray}
\mathcal{L}_\mathrm{int} &=&
 G_D \left[(q^T C i\gamma_5 
 \varepsilon\varepsilon q) 
 (\bar q i \gamma_5 C\varepsilon\varepsilon\bar q^T)
 +  (q^T C  \varepsilon\varepsilon q) 
 (\bar q  C   \varepsilon\varepsilon\bar  q^T)\right] 
\nonumber \\
&&+ \zeta G_D   \left[
(q^T C i\gamma_5 \mathcal{P}_L \varepsilon\varepsilon q)
(\bar q   i\gamma_5  \mathcal{P}_L C \varepsilon\varepsilon
\bar  q^T) + 
(q^T C i\gamma_5 \mathcal{P}_R \varepsilon\varepsilon
q) 
(\bar q i \gamma_5 \mathcal{P}_R C \varepsilon\varepsilon
\bar q^T)\right],
\label{eq:musicastranacontopi_aaammm}
\end{eqnarray}
The second line in~\eqref{eq:musicastranacontopi_aaammm}
breaks $U(1)_A$: we consider it as an effective way to model the
quark-quark interaction arising from the one-instanton exchange.

In order to couple the axions to the quarks, we note that the former
can only couple to the $U(1)_A$-breaking term in Eq.~\eqref{eq:musicastranacontopi_aaammm}
(in fact, the QCD axion couples to instanton-like gluon configurations).
In agreement with what has been done for the coupling
to quarks in the vacuum~\cite{Zhang:2023lij},
we write this coupling as
\begin{eqnarray}
\mathcal{L}_\mathrm{int} &=&
 G_D \left[(q^T C i\gamma_5 
 \varepsilon\varepsilon q) 
 (\bar q i \gamma_5 C\varepsilon\varepsilon\bar q^T)
 +  (q^T C  \varepsilon\varepsilon q) 
 (\bar q  C   \varepsilon\varepsilon\bar  q^T) \right]
\nonumber \\
&&+ \zeta G_D \left[e^{i\frac{a}{f_a}} 
(q^T C i\gamma_5 \mathcal{P}_L \varepsilon\varepsilon q)
(\bar q   i\gamma_5  \mathcal{P}_L C \varepsilon\varepsilon
\bar  q^T)+  e^{-i\frac{a}{f_a}} 
(q^T C i\gamma_5 \mathcal{P}_R \varepsilon\varepsilon
q) 
(\bar q i \gamma_5 \mathcal{P}_R C \varepsilon\varepsilon
\bar q^T)\right].
\label{eq:musicastranacontopi}
\end{eqnarray}
The Lagrangian density~\eqref{eq:musicastranacontopi} specifies the
interaction we adopt in our model.
As emphasized above, the first line 
in the right hand side of~\eqref{eq:musicastranacontopi}
preserves the $U(1)_A$ symmetry,
representing an effective way of writing the attractive
channel of the quark-quark interaction
arising from one-gluon exchange.
The axial symmetry instead is broken
by the terms in the second line \cite{tHooft:1976snw,tHooft:1986ooh}.
We note that in~\cite{Blaschke:2005uj},
authors included both terms in the
first line of~\eqref{eq:musicastranacontopi};
however, they usually
neglect the second addendum in the first line of~\eqref{eq:musicastranacontopi},
since in the mean-field, it gives rise to 
a pseudo-scalar condensate that is usually neglected:
as a matter of fact the instanton-induced interaction,
namely the second line of~\eqref{eq:musicastranacontopi},
favors condensation in the scalar channel when $a=0$
(we explicitly verified this statement within our model, see below).
However, when $a\neq 0$ condensation can happen in the pseudo-scalar
channel as well, therefore we need to consider the whole 
interaction~\eqref{eq:musicastranacontopi}.

In this work, we treat the interaction~\eqref{eq:musicastranacontopi}
within the mean field approximation.
In order to implement this, we introduce the condensates
\begin{eqnarray}
&&\langle
q^T C  i\gamma_5 \mathcal{P}_L \varepsilon\varepsilon
q
\rangle    = - h_L, \label{eq:zeta_5a}\\
&&\langle
\bar q i \gamma_5 \mathcal{P}_L C\varepsilon\varepsilon
\bar q^T
\rangle    = h_R^*, \label{eq:zeta_6a}\\
&&\langle
q^T C  i\gamma_5 \mathcal{P}_R \varepsilon\varepsilon
q
\rangle    = h_R, \label{eq:zeta_7a}\\
&&\langle
\bar q i \gamma_5 \mathcal{P}_R C\varepsilon\varepsilon\bar q^T
\rangle    = -h_L^*, \label{eq:zeta_8a}
\end{eqnarray}
as well as their combinations
\begin{eqnarray}
&&\langle
q^T C i\gamma_5 \varepsilon\varepsilon q
\rangle    = h_R - h_L, \label{eq:zeta_1a}\\
&&\langle
\bar q i \gamma_5 C\varepsilon\varepsilon\bar q^T
\rangle    = h_R^* - h_L^*. \label{eq:zeta_2a}
\end{eqnarray}
Using these,
as well as the assumptions $h_L=h_L^*$, $h_R=h_R^*$,
we get, within the mean-field approximation,
\begin{eqnarray}
\mathcal{L}_\mathrm{int} &=&
 G_D  (h_R-h_L)
\left[(\bar q i \gamma_5 C\varepsilon\varepsilon\bar q^T) 
+(q^T C i\gamma_5 \varepsilon\varepsilon q) \right]
- G_D(h_R-h_L)^2 
\nonumber \\
&&- G_D 
\left[(h_R+h_L)(\bar q i  C\varepsilon\varepsilon \bar q^T) 
-(h_R + h_L)(q^Ti C \varepsilon\varepsilon q) \right]
- G_D(h_R+h_L)^2
\nonumber \\
&&+ \zeta G_D  e^{i\frac{a}{f_a}} 
\left[
-h_L(\bar q   i\gamma_5  \mathcal{P}_L C\varepsilon\varepsilon
\bar  q^T) +h_R
(q^T C i\gamma_5 \mathcal{P}_L \varepsilon\varepsilon q) 
\right] + \zeta G_De^{i\frac{a}{f_a}} h_L h_R\nonumber\\
&&+ \zeta G_D  e^{-i\frac{a}{f_a}} 
\left[
h_R 
(\bar q i \gamma_5 \mathcal{P}_R C\varepsilon\varepsilon
\bar q^T)
-h_L(q^T C i\gamma_5 \mathcal{P}_R \varepsilon\varepsilon q) 
\right] + \zeta G_D e^{-i\frac{a}{f_a}} h_L h_R.
\label{eq:quandolavoravatenellufficiodelprocuratore}
\end{eqnarray}
Note that $\mathcal{L}_\mathrm{int} = \mathcal{L}_\mathrm{int}^\dagger$.

In order to simplify the notation we introduce
the Nambu-Gorkov bi-spinors
 \begin{equation}
     \Psi=\left(\begin{array}{c}
            q\\
           C \bar q^T
     \end{array}\right), \qquad \bar \Psi= \left(\bar q, q^T C\right).
 \end{equation}
In terms of these, we then can rewrite the
interaction lagrangian density
as
\begin{equation}%\label{eq:lintfin}
     \mathcal{L}_\mathrm{int}=  
     \mathcal{V}
     + \bar \Psi \bm\Delta \Psi,
\label{eq:mapoiimmaginainiziaabatterelepalpebre}
 \end{equation}
where 
\begin{equation}
\mathcal{V}=- 2G_D(h_R^2+h_L^2) +2\zeta G_D h_L h_R \cos\left(
\frac{a}{f_a}\right),\label{eq:potenzialeV_abc}
\end{equation}
and
\begin{equation}
\bm\Delta=\left(
     \begin{array}{cc}
         0 & \Phi^- \\
         &\\
       \Phi^+  &  0
     \end{array}\right),
\label{eq:anterkaskow}
 \end{equation}
with
\begin{eqnarray} \label{eq:coeff}
\Phi^- &=& G_D\left[ 2 h_R\mathcal{P}_L-2 h_L\mathcal{P}_R
    - e^{ia/f_a} h_L\zeta   \mathcal{P}_L
    + e^{-ia/f_a}h_R\zeta    \mathcal{P}_R\right]
    i\gamma_5\varepsilon_{ij}\varepsilon_{\alpha\beta 3}
    \\
\Phi^+ &=& G_D\left[2 h_R \mathcal{P}_R-2 h_L\mathcal{P}_L  + e^{ia/f_a}h_R \zeta   \mathcal{P}_L
    -e^{-ia/f_a}h_L \zeta    \mathcal{P}_R\right] 
    i\gamma_5\varepsilon_{ij}\varepsilon_{\alpha\beta 3}.
    \end{eqnarray}
In the above equations, the first two terms 
 arise from the one-gluon exchange interaction,
while those proportional to $\zeta$ arise from the
one-instanton exchange. 
The matrix $\bm\Delta$ has a similar structure to that of~\cite{Ruester:2005jc}:
indeed, it satisfies $\Phi^+  = \gamma_0 (\Phi^-)^\dagger \gamma_0$.

To the interaction~\eqref{eq:mapoiimmaginainiziaabatterelepalpebre} we
need to add the kinetic term of quarks
at finite chemical potential $\mu$.
This contribution is well-known~\cite{Blaschke:2005uj,Ruester:2005jc} 
and leads to the
full lagrangian, that in momentum space reads
\begin{equation}
    \mathcal{L} =  \mathcal{V} + 
      \bar\Psi S^{-1} \Psi.
 \end{equation}
Here, the inverse
quark propagator is given by
\begin{equation}
     S^{-1}(p)=\left(
     \begin{array}{cc}
        (\slashed{p}+ \mu \gamma_0)\bm 1_C\bm 1_F \ & \Phi^-  \\
         &\\
       \Phi^+ &  (\slashed{p}- \mu \gamma_0)\bm 1_C\bm 1_F
     \end{array}\right),
\end{equation}
where $\bm 1_C$ and $\bm 1_F$ correspond to the identities
in color and flavor spaces respectively. 
We note that these matrices have dimension $4\times 3\times 2 \times 2=48$, due to Dirac, color, flavor, and Gorkov indices, respectively.

\subsection{Thermodynamic potential\label{sec:ld}}

The thermodynamic potential is obtained via the 
standard integration over the fermion fields in 
the partition function~\cite{Blaschke:2005uj,Ruester:2005jc}, which leads at 
\begin{equation}
    \Omega=-\mathcal{V}+\Omega_{1-\mathrm{loop}},
\end{equation}
where 
$\mathcal{V}$ denotes the mean field contribution~\eqref{eq:potenzialeV_abc}, and
$\Omega_{1-\mathrm{loop}}$ corresponds to the 1-loop contribution of the quarks, namely
\begin{equation}
  \Omega_{1-\mathrm{loop}}=-T\sum_n\int \frac{d^3 \,p}{(2\pi)^3} \frac{1}{2}
  \mathrm{Tr}
  \log  \left(\beta S^{-1}(i\omega_n, \vec{p})\right). 
  \label{eq:musicaerumoredibottiglie}
\end{equation}
In getting Eq.~\eqref{eq:musicaerumoredibottiglie} we
performed the functional integral over fermions 
at fixed $a$: in fact, in this work the axion field
is treated as a classical background.
Consequently, the thermodynamic potential is a function
of $a$ and physical quantities have to be computed
at a given value of $a$.
Moreover,
we used the 
imaginary time formalism of finite temperature field theory,
and $\omega_n=(2n+1)\pi T $ are the relevant 
fermionic Matsubara frequencies.
The overall $1/2$ 
in~\eqref{eq:musicaerumoredibottiglie} 
takes into account the artificial doubling of
degrees of freedom introduced by shifting to the
Nambu-Gorkov bi-spinors. Finally,
the trace is understood in 
Nambu-Gorkov, Dirac, color and flavor spaces.

In order to evaluate the sum
over $\omega_n$ in 
Eq.~\eqref{eq:musicaerumoredibottiglie}, 
we follow the well-known strategy of computing
the eigenvalues of the matrix~\cite{Blaschke:2005uj,Ruester:2005jc}
\begin{equation}
 \mathcal{T}=\left(
 \begin{array}{cc}
(-\gamma_0 \vec{p}\cdot\vec{\gamma}+ \mu)\bm 1_C\bm 1_F 
\ & \gamma_0 \Phi^-  \\
         &\\
      \gamma_0 \Phi^+  &  (-\gamma_0\vec{p}\cdot \vec{\gamma}- \mu )\bm 1_C\bm 1_F 
     \end{array}\right).
\end{equation}
These are given by
\begin{eqnarray}
   \varepsilon_{1,\pm} &=&\pm|p-\mu|,\label{eq:iprimi}\\
      \varepsilon_{2,\pm} &=&\pm|p+\mu|,\label{eq:iSecondi}\\
     \varepsilon_{3,\pm} & =&\pm \sqrt{(p - \mu)^2+ \Delta_3^2},\\
     \varepsilon_{4,\pm} & =&\pm \sqrt{(p + \mu)^2+ \Delta_3^2},\\
 \varepsilon_{5,\pm}&=&\pm \sqrt{(p - \mu)^2+ \Delta_5^2},\\
 \varepsilon_{6,\pm}&=&\pm \sqrt{(p + \mu)^2+ \Delta_5^2},
\end{eqnarray}
where 
$p\equiv|\bm p|$ denotes the magnitude of the $3-$momentum $\bm p$;
each of the eigenvalues above has a multiplicity equal to $4$. 
The eigenvalues~\eqref{eq:iprimi} and~\eqref{eq:iSecondi}
correspond to those of the blue quarks, which do not participate
to the pairing and remain ungapped.
Moreover, we put
\begin{eqnarray}
\Delta_3^2 &=&\zeta^2 G_D^2 h_L^2 + 4 G_D^2 h_R^2 - 4 \zeta G_D^2 h_L  h_R \cos(a/fa),\label{eq:delta_3_squa}\\
\Delta_5^2 &=& \zeta^2 G_D^2 h_R^2  + 4 G_D^2 h_L^2 - 4 \zeta G_D^2 h_L h_R  \cos(a/fa),
\label{eq:delta_5_squa}
\end{eqnarray}
that represent the (squared) gaps in the quark spectrum.

Putting $\varepsilon_k(\vec{p})=\varepsilon_{k,+}(\vec{p})$, $k=1,\cdots, 6$, we use 
$\ln\det \beta S^{-1}= \mbox{Tr}\ln  \beta S^{-1}$ and write 
\begin{equation}
    \ln\det \beta S^{-1}_L(i\omega_n, \vec{p})= 4\sum_{k=1}^{6} \ln \left(\frac{\omega_n^2+\varepsilon_k(\vec{p})^2}{T^2}\right),
\end{equation}
where we made explicit the degeneracy, 
equal to $4$,
of the eigenvalues and matched the ones with opposite signs.
Now the Matsubara sum can be easily evaluated, since
\begin{equation}
    T\sum_n \ln\left(\frac{\omega_n^2+\varepsilon_k(\vec{p})^2}{T^2}\right) = |\varepsilon_k(\vec{p})|+ 2T \ln \left(1+e^{-|\varepsilon_k(\vec{p})|/T}\right),
\end{equation}
so the 1-loop contribution finally results in 
\begin{equation}
\Omega_{1-\mathrm{loop}}^{L}=-2\sum_{k=1}^{6}\int \frac{d^3p}{(2\pi)^3} \left[\varepsilon_k(\vec{p})+ 2T \ln \left(1+e^{-\varepsilon_k(\vec{p})/T}\right)\right],
 \label{eq:the_one_loop_uua}
\end{equation}   
where we took into account that according to our definitions
the $\varepsilon_k$ are always positive.
Therefore, the thermodynamic potential is given by
\begin{equation}
\Omega=2G_D(h_R^2+h_L^2) -2\zeta G_D h_L h_R \cos\left(
\frac{a}{f_a}\right)
-2\sum_{k=1}^{6}\int \frac{d^3p}{(2\pi)^3} \left[\varepsilon_k(\vec{p})+ 2T \ln \left(1+e^{-\varepsilon_k(\vec{p})/T}\right)\right].
\label{eq:Omega_kkk_ll_ppp}
\end{equation}

The first addendum in the right-hand side of~\eqref{eq:the_one_loop_uua}
corresponds to the $T=0$ contribution to the thermodynamic
potential, which is divergent in the ultraviolet. In order to
regulate this divergence, we introduce a sharp 3D momentum cutoff,
$\Lambda$, and rewrite that term as
\begin{equation}
\int \frac{d^3p}{(2\pi)^3} \varepsilon_k(p)=
\frac{4\pi}{8\pi^3}\int_0^\Lambda p^2dp ~\varepsilon_k(p).
\label{eq:the_vac_term_LA}
\end{equation}
The cutoff roughly represents the momentum scale above which
the contact interaction used in the present work should replaced by
a non-local term directly borrowed from QCD. 
We treat $\Lambda$ as a free parameter of the model,
and from previous studies based on the NJL model, we assume
$\Lambda=O(1~\mathrm{GeV})$. {We also notice that we inserted no cutoff in the thermal part because it is finite; however in previous studies, based on NJL-models at zero baryon density, this cutoff was  introduced and it was shown to mildly affect thermodynamic quantities, particularly near the critical temperature for chiral-symmetry restoration \cite{Ratti:2005jh,Ratti:2004ra,Ruggieri:2016ejz}.
In fact, the cut of high momenta in the loop could be interpreted as a very rough implementation of a momentum-dependent  mass function. The analysis of this occurence, although very interesting, is beyond the scope of the present study, therefore we leave  it to a future work.}

We notice that although in~\eqref{eq:the_one_loop_uua} we sum over
the six positive eigenvalues of $\mathcal{T}$, the ones corresponding
to the blue quarks, which do not participate in the pairing,
do not contribute to the value of the condensates.

It is useful to stress that both $\mathcal{V}$ and the quark
spectrum
are invariant under the set of transformations
$h_L \leftrightarrow  h_R$ and
$h_L \leftrightarrow  -h_R$; moreover,
if $\zeta=0$ then $\Omega$\
depends on $h_L^2$ and $h_R^2$ only: as a consequence, in this limit
we expect $\Omega$ to develop a set of four degenerate minima
along the lines $h_L = \pm h_R$. 
This degeneracy is removed by $\zeta\neq 0$.
This picture is confirmed by the direct
evaluation of $\Omega$, see Fig.~\ref{Fig:potential_comparison_senzatheta}
in the next section.

\subsection{Thermodynamic potential at finite $\theta=a/f_a$:
the lines $h_L = \pm h_R$}

\begin{figure*}[t!]
    \centering
    \includegraphics[scale=0.4]{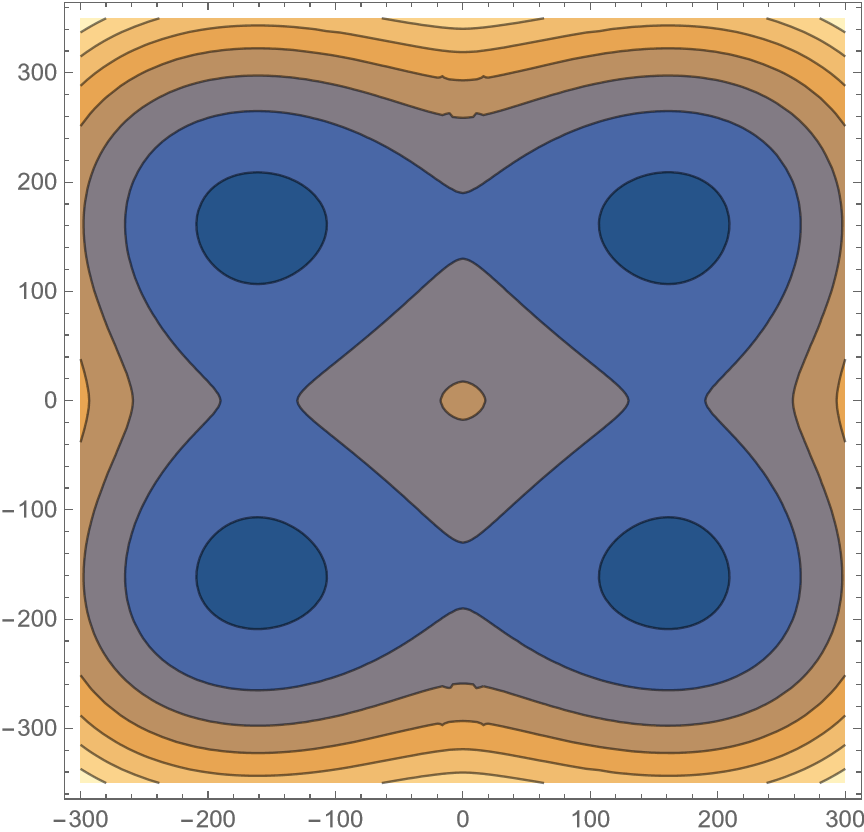}~~~
    \includegraphics[scale=0.4]{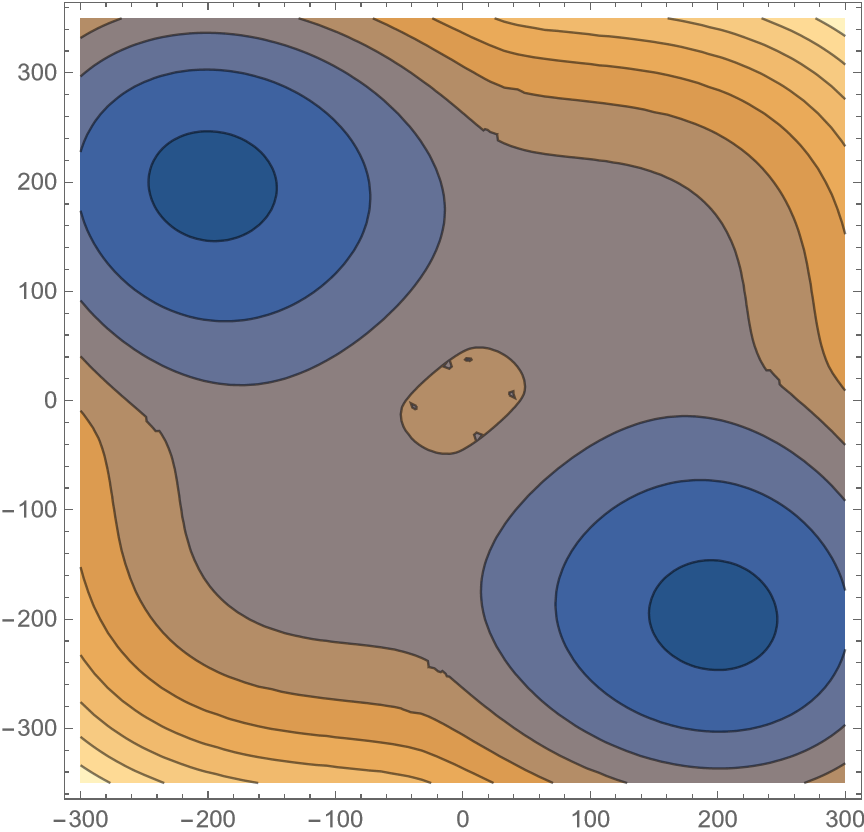}
    \caption{Thermodynamic potential at $T=0$, computed for 
    $\theta\equiv a/f_a=0$,
  and for  
    $\zeta=0$ (left panel) and for  
    $\zeta=0.2$ (right panel)
    In the plots, $x$ and $y$-axes denote $\Delta_L$ and $\Delta_R$
    respectively (measured in MeV). }
    \label{Fig:potential_comparison_senzatheta}
\end{figure*}

In this subsection, we analyze the shape of $\Omega$
in the $(h_L,h_R)$ plane: this is preparatory to the results
on the gap that we present in the next section.
In Fig.~\ref{Fig:potential_comparison_senzatheta}
we plot the thermodynamic potential at $T=0$, computed for 
    $\theta\equiv a/f_a=0$,
  and for  
    $\zeta=0$ (left panel) and for  
    $\zeta=0.2$ (right panel).
    In the plots, $x$ and $y$-axes denote $\Delta_L$ and $\Delta_R$
    respectively (measured in MeV),
    which are defined as
\begin{equation}
\Delta_L = 2G_D h_L,~~~\Delta_R = 2 G_D h_R.
\label{eq:def_capital_DDD_aa}
\end{equation}
It is more convenient to use $\Delta_{L,R}$ instead of
$h_{L,R}$ since the former 
correspond to the gaps in the quark spectrum when $\zeta=0$.
From these, {we can introduce the following scalar and pseudo-scalar combinations} as
\begin{equation}
\Delta_\mathrm{S} = \Delta_R - \Delta_L,~~~\Delta_\mathrm{PS}=\Delta_R + \Delta_L.\label{eq:cheprocuralautogoal}
\end{equation}
{We notice in Fig. 1 that for $\zeta$ = 0, namely when we only consider the
one-gluon-exchange interaction, there are four loci of degenerate minima. Two of
these correspond to $\Delta_L = \Delta_R$ and therefore to finite pseudoscalar condensate and
vanishing scalar condensate whereas the other two correspond to $\Delta_L = \Delta_R$ and
consequently to finite scalar condensate and vanishing pseudoscalar condensate}.

On the other hand,
when the instanton-induced interaction is switched on $(\zeta\neq0)$
this degeneracy is removed, and the condensation in the 
scalar channel $\Delta_L = -\Delta_R$ is favored.

\begin{figure*}[t!]
    \centering
    \includegraphics[scale=0.4]{plot_b_0dot2_a_0.png}~~~
    \includegraphics[scale=0.4]{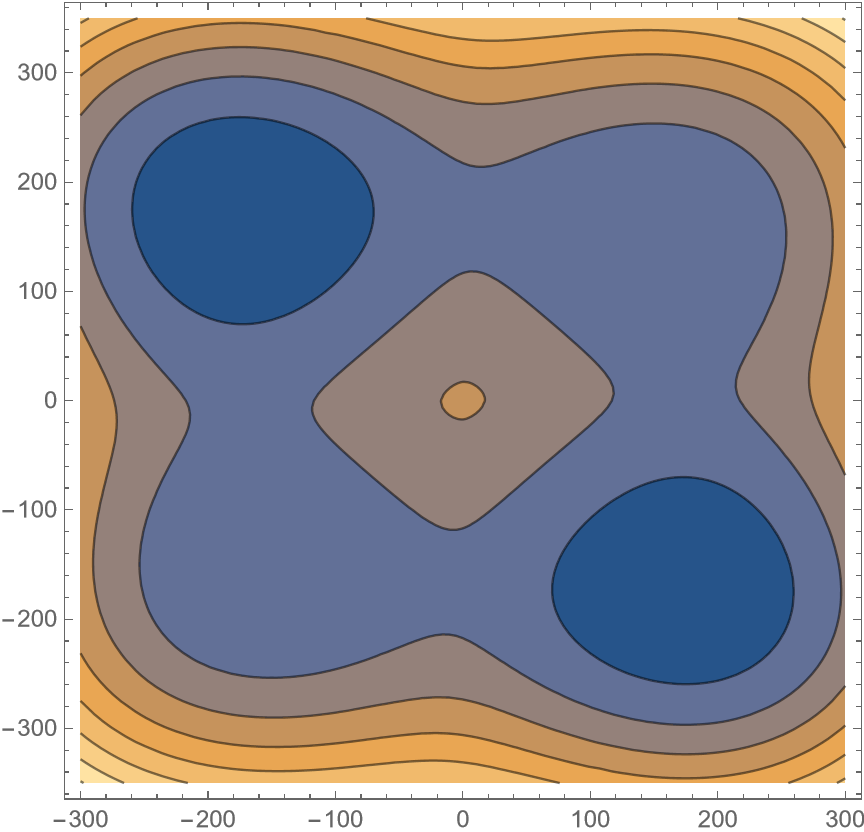}\\
    \includegraphics[scale=0.4]{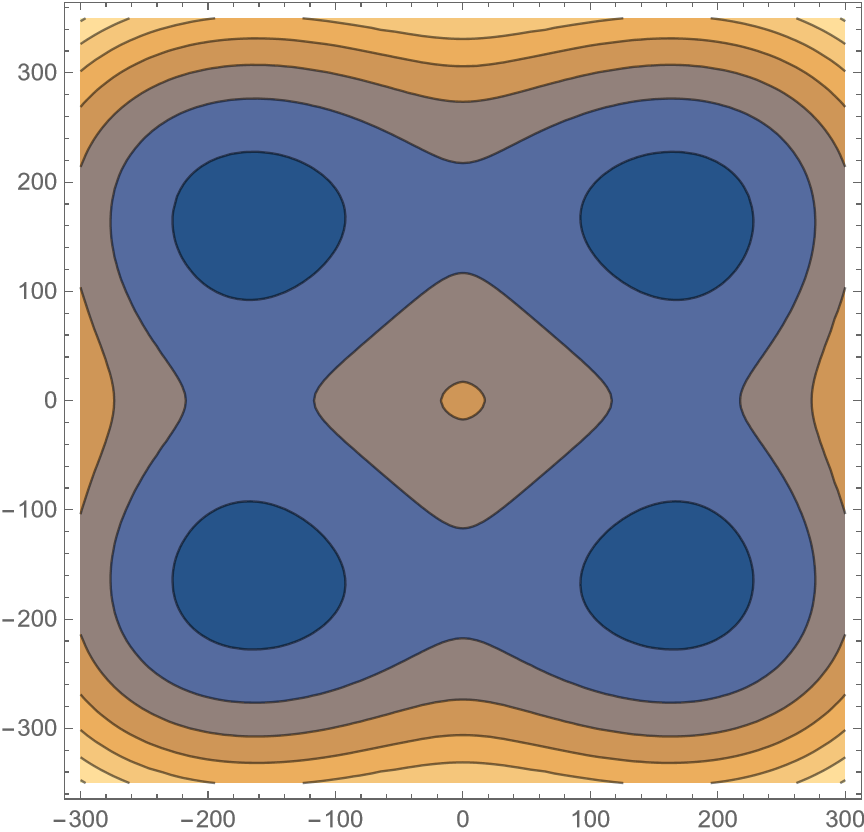}~~~
    \includegraphics[scale=0.4]{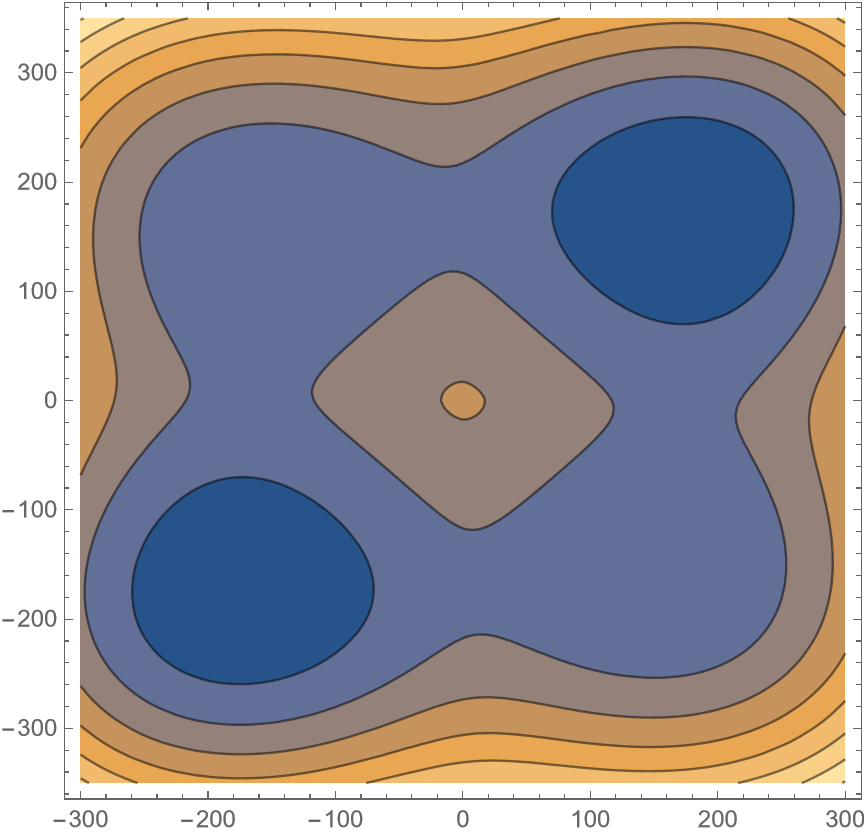}\\
    \includegraphics[scale=0.4]{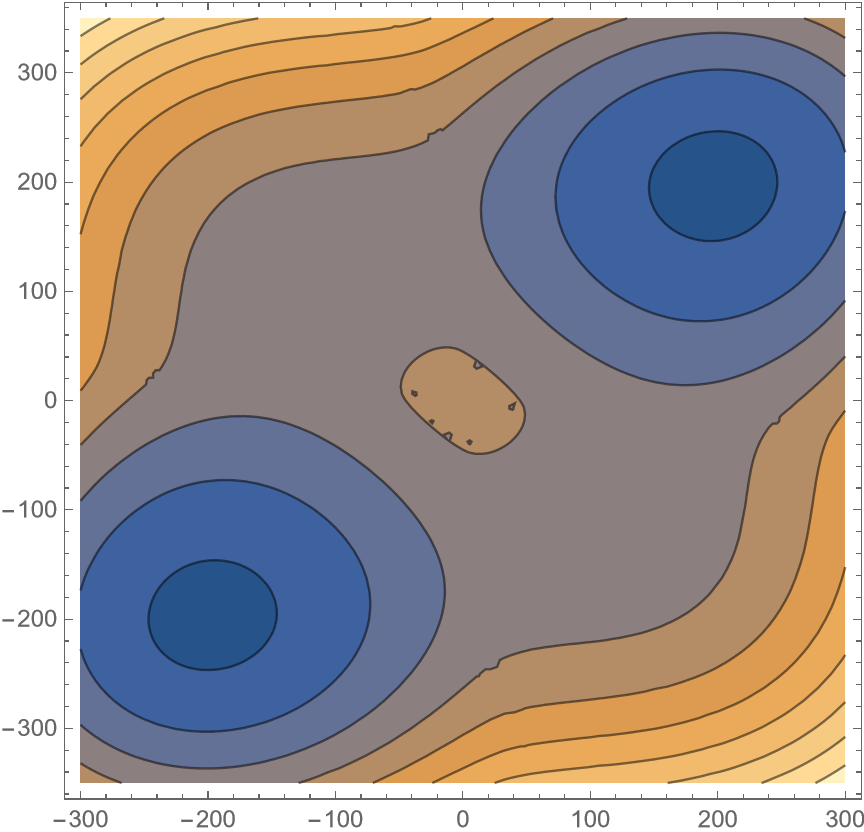}~~~
    \includegraphics[scale=0.4]{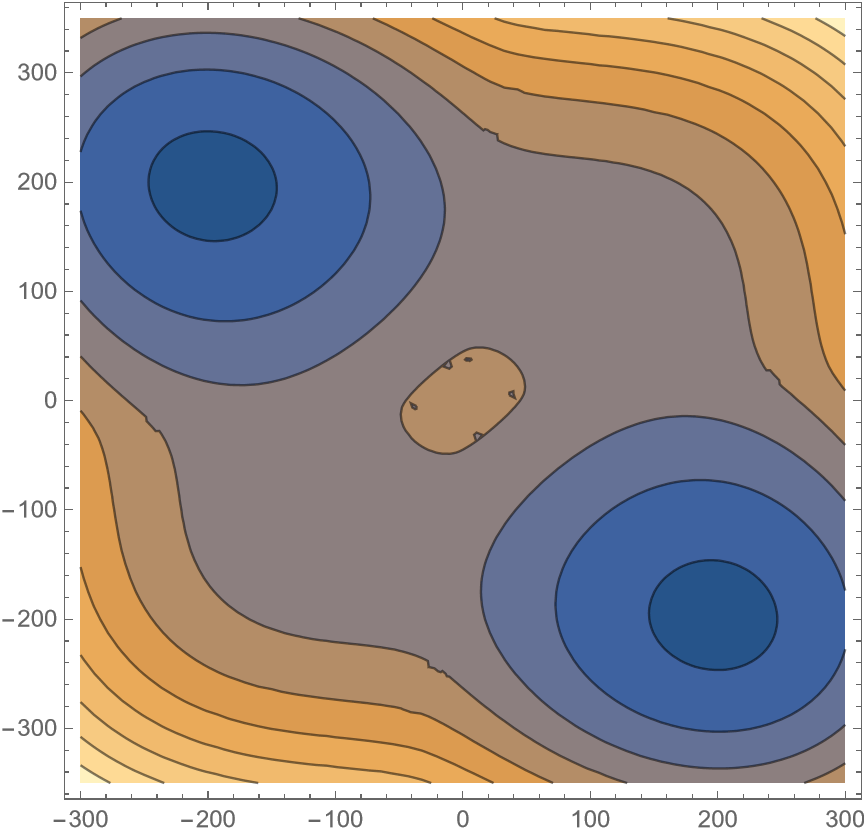}
    \caption{Thermodynamic potential at $T=0$, computed for $a_1=a_2=1$,
    $\zeta=0.2$, and several values of $\theta\equiv a/f_a$. 
    In each plot, $x$ and $y$-axes denote $\Delta_L$ and $\Delta_R$
    respectively (measured in MeV). Top-left plot corresponds to
    $\theta=0$, top-right to $\theta=\pi/2-\varepsilon$,
    center-left to $\theta=\pi/2$, center-right to $\theta=\pi/2+\varepsilon$, {with $\varepsilon=0.25$}, bottom-left to $\theta=\pi$,
    and finally bottom-right to $\theta=2\pi$.}
    \label{Fig:potential_comparison}
\end{figure*}

In Fig.~\ref{Fig:potential_comparison}
we plot the
thermodynamic potential at $T=0$, computed for 
    $\zeta=0.2$, and several values of $\theta\equiv a/f_a$. 
    In each plot, $x$ and $y$-axes denote $\Delta_L$ and $\Delta_R$
    respectively (measured in MeV).
In the figure, the top-left plot corresponds to
$\theta=0$, top-right to $\theta=\pi/2-\varepsilon$,
center-left to $\theta=\pi/2$, center-right to $\theta=\pi/2+\varepsilon$,  {with $\varepsilon=0.25$}, bottom-left to $\theta=3\pi/2$,
and finally bottom-right to $\theta=2\pi$.
For all the cases shown in the figure, 
the minima of $\Omega$ 
sit on the lines $\Delta_L = \pm \Delta_R$:
this is the obvious consequence of the fact that
$\Omega$ is invariant under the transformations
$h_L \leftrightarrow  \pm h_R$.
For $a/f_a$ in the range $[0,\pi/2)$, 
the global minima correspond to $\Delta_L = - \Delta_R$;
hence, in this range of $a/f_a$ only the scalar condensate exists.
The minima {become less shallow} as 
$a/f_a$ is increased: indeed, 
for $a/f_a$ in the range $(\pi/2, 3\pi/2)$ the global minima
are located along the line $\Delta_L = \Delta_R$, implying 
a phase transition from the scalar to the
pseudo-scalar condensate.
Finally, increasing $a/f_a$ up to $2\pi$ the location
of the minima changes again, resulting in another phase transition
from the pseudo-scalar to the scalar condensate.
Along the lines $\Delta_L =\pm \Delta_R$ we have  
$\Delta_3=\Delta_5$
for any $a$ and $\zeta$:
hence, there is only one gap in the quark spectrum,
see~\eqref{eq:delta_3_squa} and~\eqref{eq:delta_5_squa}.

We notice that the results on the minima of $\Omega$ discussed above
stand for $\zeta>0$: we checked that for $\zeta<0$ the role 
of the scalar and the pseudoscalar condensates invert;
besides this, there is no major difference between the system with
positive and negative $\zeta$. Consequently, from now on we limit 
ourselves to show results for $\zeta>0$ only.

Before going ahead, 
we determine the allowed values of $\zeta$
that lead to the nontrivial solution of the gap equation.
To this end, it is enough to limit ourselves to the gap equation
at $T=0$ and small $\theta$.
Imposing the stationarity condition of $\Omega$ with respect to $h_L$,
namely $\partial\Omega/\partial h_L=0$ with $\Omega$
given by Eq.~\eqref{eq:Omega_kkk_ll_ppp},
{and then applying this condition along the line}
$h_L = - h_R$, 
which is the one along which $\Omega$ develops minima
for small $\theta$ as discussed above,
gives
\begin{equation}
2 + \zeta \cos(a/f_a) = 
\frac{G_D}{2\pi^2}\int_0^\Lambda p^2dp~
\left(
\frac{A(a,\zeta)}{\sqrt{(p-\mu)^2 + G_D^2 h_L^2 A(a,\zeta)}}
+\mu\rightarrow-\mu\right),
\label{eq:matchingorganicmatter}
\end{equation}
where we removed the trivial solution $h_L=0$, and 
we defined
\begin{equation}
A(a,\zeta) =  4 + 4\zeta\cos(a/f_a) + \zeta^2.
\label{eq:thosearethefossilsofmicrobes}
\end{equation}
The above equation shows that
the condition $ \zeta <2$ must be satisfied
in order to have a non-trivial solution of the gap equation.
In fact, the right-hand side
of~\eqref{eq:matchingorganicmatter} is always positive, and so
the left-hand side must be positive as well: 
{this can be obtained for all
values of $a$ if and only if $\zeta<2$.}
For $ \zeta \ge 2$ there is a range of $a$ in which the gap equation
has only the trivial solution $h_L=0$: hence, in what follows
we limit ourselves to $ \zeta <2$.

%} 

\section{Results}

\subsection{Gaps versus $a/f_a$}

\begin{figure}[t!]
    \centering
    \includegraphics[scale=0.18]{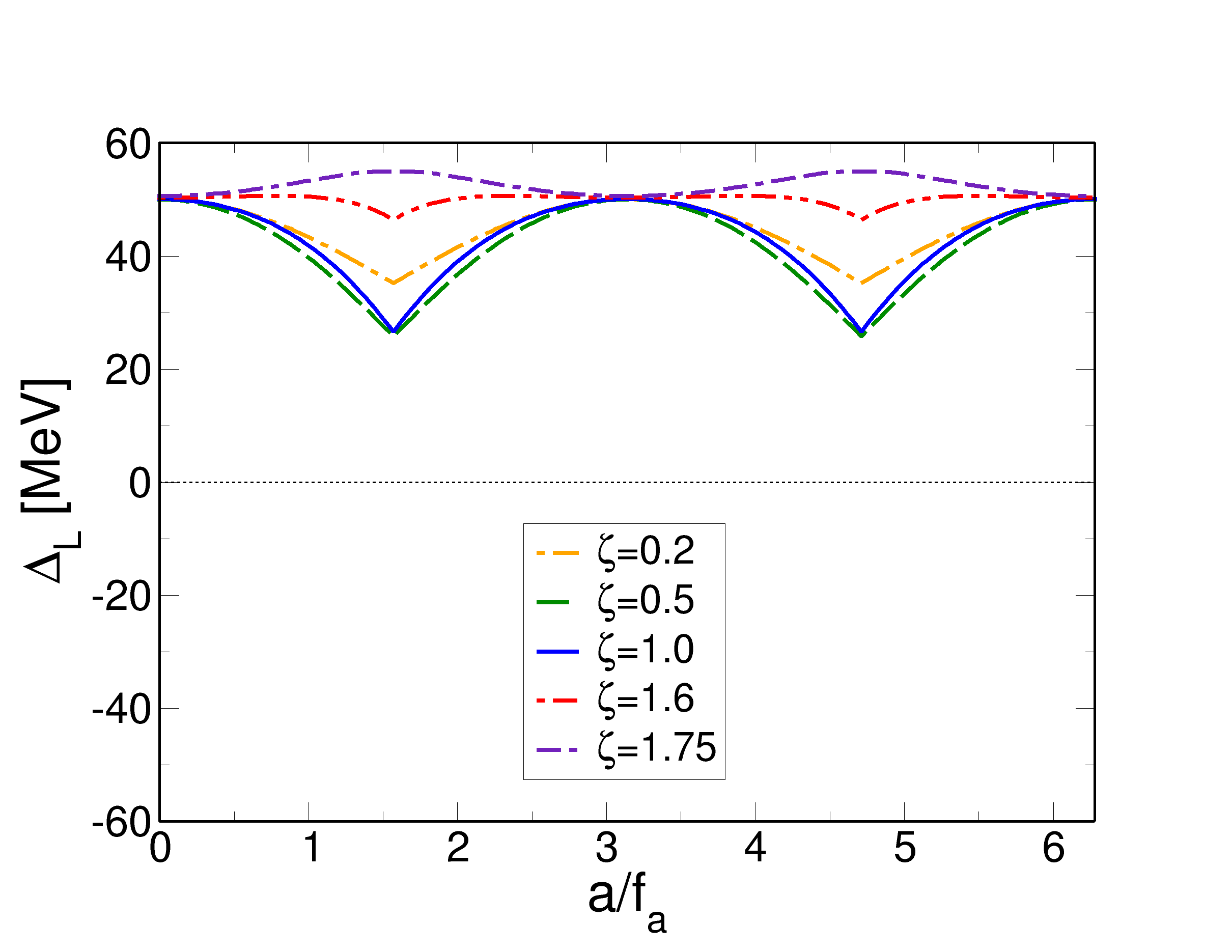}~~~
    \includegraphics[scale=0.18]{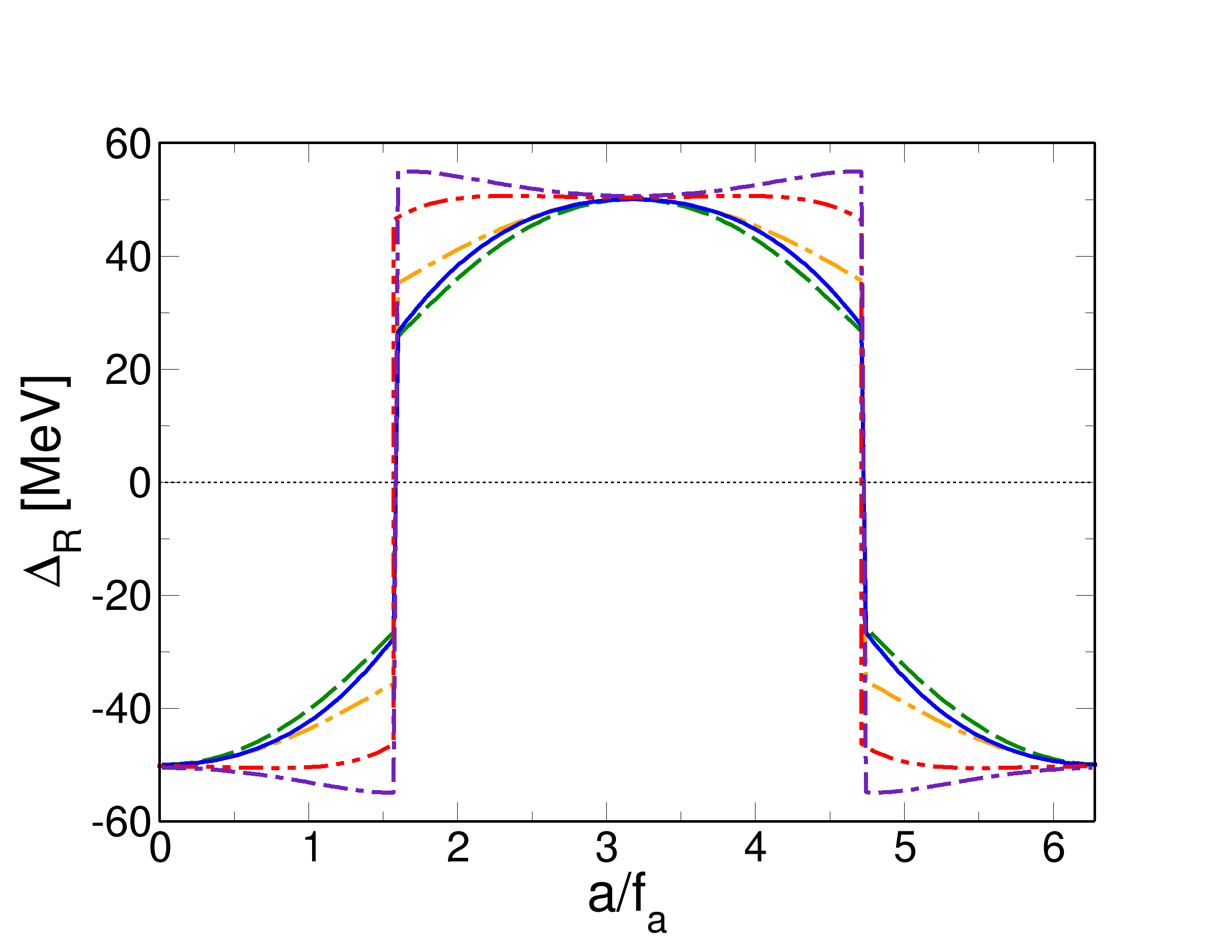}
    \caption{Gaps $\Delta_L$ and $\Delta_R$  at 
    $T=0$ and $\mu=400$ MeV, 
    versus $a/f_a$,
    and for several values of $\zeta$.
    $G_D$ is fixed in order to have $\Delta_L=50$ MeV
    for $a=0$ at $\mu=400$ MeV.}
    \label{Fig:gaps}
\end{figure}

In this subsection, we present our results for the gap 
parameters.
For $\Delta_L = \pm \Delta_R$,
which correspond to the directions along which
the minima develop at finite $a$,
we get 
from Eqs.~\eqref{eq:delta_3_squa} and~\eqref{eq:delta_5_squa}
that $\Delta_3 = \Delta_5$.
Therefore we limit ourselves to show results
for $\Delta_3$ only.
We fix $\mu=400$ MeV and $\Lambda=1$ GeV
as representative values
of the quark chemical potential and the UV cutoff.
For each $\zeta$, we fix $G_D$ in order to have a desired
value of $\Delta_L$ at $a=0$.

In Fig.~\ref{Fig:gaps} we plot 
$\Delta_L$ and $\Delta_R$ 
versus $a/f_a$ at $T=0$ and $\mu=400$ MeV;
we consider several values of $\zeta$,
while
$G_D$ is fixed for each $\zeta$ so that $\Delta_L=50$ MeV
at $a=0$.
We firstly focus on the results for $\zeta=0.2$, $0.5$ and $1$.
For these values of $\zeta$, 
we notice that 
for $a/f_a$ in the range $(0,\pi/2)$,
$\Delta_R = - \Delta_L$, hence only the scalar condensate forms.
On the other hand, for $a/f_a$ in the range $(\pi/2,3\pi/2)$,  
$\Delta_R =  \Delta_L$. In this case,
only the pseudo-scalar condensate forms.
Finally,  for 
$a/f_a$ in the range $(3\pi/2,2\pi)$ we find again
$\Delta_R =  -\Delta_L$, hence 
there is a phase transition from the pseudo-scalar to the
scalar condensate.
We also notice that for {$\zeta=1.6$ and} $\zeta=1.75$
the magnitude of $\Delta_L$ increases with
$a$ for $a/f_a$ in the range
$(0,\pi/2)$. 

{The results in Fig.~\ref{Fig:gaps}
show several interesting features.
In fact, as we already pointed out, there is a noticeable qualitative difference in dependency of $\Delta_L$ when comparing the three lowest values of $\zeta$ with the higher ones. In particular, $\Delta_L$ decreases close to $a/f_a=0$ for $\zeta=0.2$,$\zeta=0.5$ and $\zeta=1.0$ and it exhibits a local minimum at $a/f_a = \pi/2$ and $3\pi/2$. Furthermore one can observe a discontinuity in its derivative at the aforementioned minimum points, for the three lowest values of $\zeta$, where the derivative changes sign.}
{In contrast, for the highest value of $\zeta$, i.e. $\zeta=1.75$, we observe that $\Delta_L$ increases in the proximity of $a/f_a=0$. }
{This suggests the presence of a critical value of $\zeta$ where the change from positive to negative curvature at $a/f_a=0$ occurs. This is indeed the case and its value is derived in Eq.(49).} 
{The change in the behaviour of $\Delta_L$ can already be observed for an intermediate value of $\zeta$, i.e., $\zeta=1.6$. In this case one can see that the curve in the proximity of $a/f_a$ is still increasing but almost flat, signaling the proximity to the aforementioned critical point. Furtermore we continue to observe local minima and  cusps at $a/f_a = \pi/2$ and $3\pi/2$. This signals the emergence of two local maxima in the ranges $a/f_a=[0,\pi/2]$ and $a/f_a=[\pi/2,\pi]$ (and the corresponding ones in the ranges  $a/f_a=[0,\pi/2]$ and $a/f_a=[\pi/2,\pi]$ considering the periodicity of $\pi$).}
{Although less evident, the same behaviour is also present for $\zeta=1.75$.}
{According to our line of reasoning then, we also conclude that $a/f_a=n\pi$ ($n$ integer) turn from being  local maxima in the case of lower values of $\zeta$, to local minima for the higher ones. }
{From our previous discussion on the transition between scalar and pseudo-scalar condensates, the behaviour of $\Delta_R$, shown in the left panel of Fig.3, can be straightforwardly understood from the corresponding one of $\Delta_L$.}

The behavior of $\Delta_L$ near $a=0$
can be semi-quantitatively understood by the approximate solution
of the gap equation~\eqref{eq:matchingorganicmatter},
namely
\begin{equation}
\Delta_L = \frac{4\delta}{\sqrt{A(a,\zeta)}}
\exp\left(
-\frac{\pi^2}{\mu^2 G_D \Theta(a,\zeta)}
\right),
\label{eq:leinonlomerita}
\end{equation}
where $A$ is given by Eq.~\eqref{eq:thosearethefossilsofmicrobes} and
\begin{equation}
\Theta(a,\zeta) = \frac{A(a,\zeta)}
{2 + \zeta \cos(a/f_a)}.
\label{eq:inunascatola}
\end{equation}
The approximated solution~\eqref{eq:leinonlomerita}
can be obtained within the 
High Density
Effective Theory (HDET) of QCD, see~\cite{Nardulli:2002ma} and 
references therein, as well as appendix~\ref{sec:appendix:hdet};
it has the standard form of the Bardeen-Cooper-Schrieffer (BCS) 
gap in the theory of
superconductivity~\cite{Bardeen:1957kj,Bardeen:1957mv}: 
in fact, $\mu^2$ is proportional to the
density of states of the pairing quarks at the Fermi surface,
and $\delta$ in~\eqref{eq:leinonlomerita}
plays the role of the Debye frequency, $\omega_D$, 
of the BCS theory, which cuts 
high momentum modes
out of the pairing. 
The solution~\eqref{eq:leinonlomerita} is strictly valid only
in the weak coupling, hence it cannot quantitatively reproduce
the results in Fig.~\ref{Fig:gaps}: however, it is still helpful
to grasp the behavior of $\Delta_L$ near $a=0$.
In fact, from~\eqref{eq:leinonlomerita} we notice that
for a fixed value of $\zeta$,
the coupling of the superconductive quarks to the axion field
effectively affects the pairing in two ways.
Firstly, it changes the width of the shell around the
Fermi surface that contributes to the pairing, 
namely
\begin{equation}
\frac{2\delta}{\sqrt{A(0,\zeta)}}\rightarrow \frac{2\delta}{\sqrt{A(a,\zeta)}}.
\label{eq:iosonomarkfrenaul}
\end{equation}
Secondly, it effectively changes the chemical potential
of the quarks, that is
\begin{equation}
\mu^2\Theta(0,\zeta)
\rightarrow
\mu^2\Theta(a,\zeta).
\label{eq:disolitosonoivecchiadareconsigli}
\end{equation}
Both $A(a,\zeta)$ and $\Theta(a,\zeta)$ are decreasing functions
of $a$ at fixed $\zeta$. Consequently, the 
response of $\Delta_L$ to the
coupling with the
axion has two competing effects: 
on the one hand, it 
leads to the opening of the shell around
the Fermi surface, thus increasing the portion of phase space
involved in the pairing; on the other hand, 
it effectively lowers the chemical potential of the paired quarks,
implying the decrease of the volume of phase space available
for pairing. Which of the two effects wins depends, at a given $G_D$,
on the value of $\zeta$. 
This can be seen by inspecting the curvature of
$\Delta_L$ for small $\theta=a/f_a$.
To this end, it is enough to expand
Eq.~\eqref{eq:leinonlomerita} near $\theta=0$,
getting
\begin{equation}
\Delta_L = \Delta_{L,0}\left(
1 + \frac{\kappa}{2}\theta^2
\right),
\label{eq:gentechegridasulbus}
\end{equation}
where $\Delta_{L,0}$ denotes the gap~\eqref{eq:leinonlomerita}
at $\theta=0$, and the curvature is
\begin{equation}
    \kappa=\zeta\frac{\pi^2 ( \zeta-2)  +2 G_D \mu^2 (2 + \zeta) }{ G_D \mu^2 (2 + \zeta)^4 }.
 \label{eq:macertocheloso}   
\end{equation}
The curvature is trivially zero for $\zeta=0$, as well as for
\begin{equation}
\zeta=\zeta_\mathrm{crit}=\frac{2(\pi^2-G_D \mu^2)}{\pi^2 + G_D \mu^2};
\label{eq:sembrachesiaunpensierocomune}
\end{equation}
it is negative for $\zeta$ in the range $(0,\zeta_\mathrm{crit})$, 
positive otherwise. Hence, the response of $\Delta_L$ to $\theta$
around $\theta=0$ depends on the specific value of $\zeta$,
and the turning point of $\zeta$,
namely $\zeta_\mathrm{crit}$, depends on the value of $G_D$.

\begin{figure}[t!]
    \centering
    \includegraphics[scale=0.18]{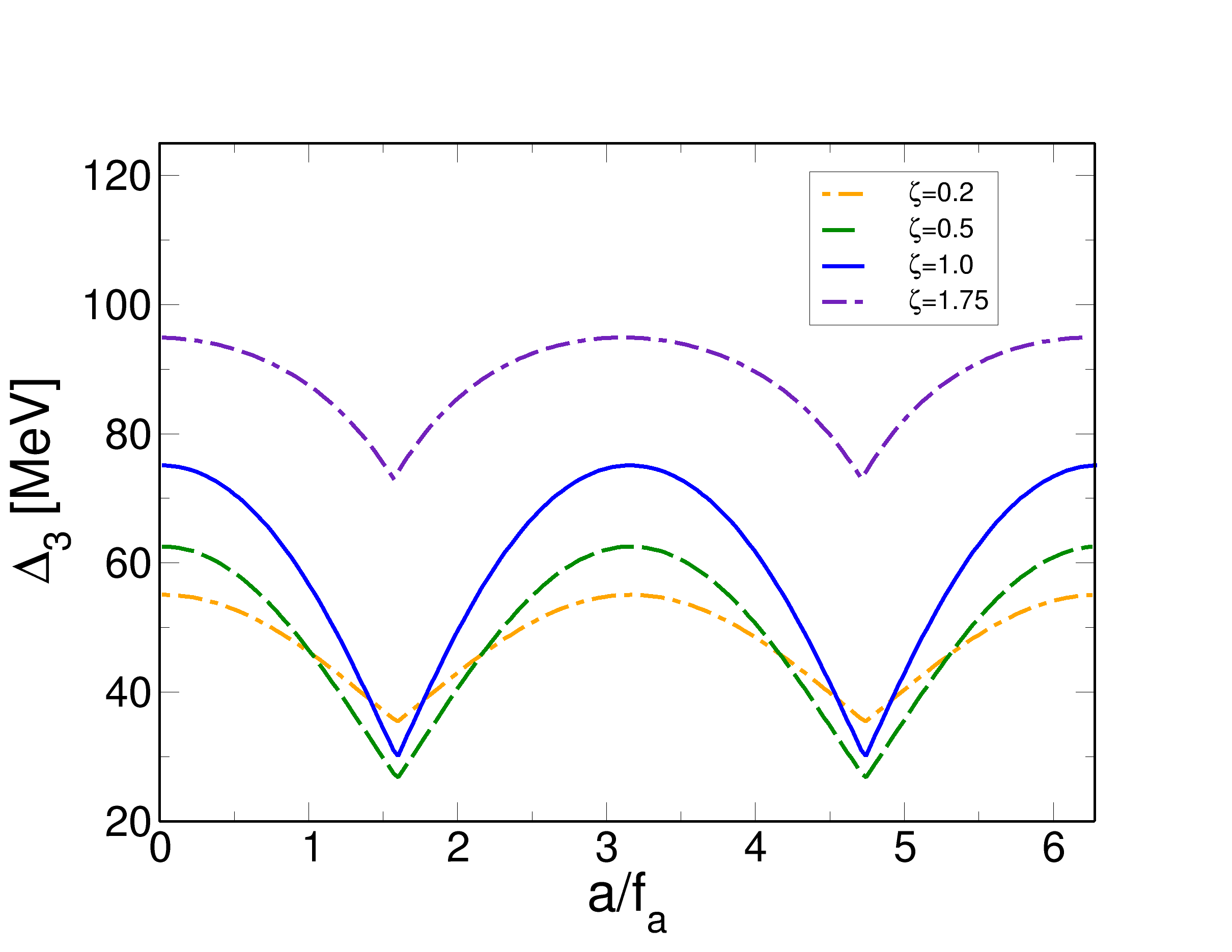}
     \caption{Gap  in the quark spectrum,
     $\Delta_3$, at 
    $T=0$ and $\mu=400$ MeV, 
    versus $a/f_a$,
    and for several values of $\zeta$.
    $G_D$ is fixed in order to have $\Delta_L=50$ MeV
    for $a=0$ at $\mu=400$ MeV.}
    \label{fig:gap3}
\end{figure}

The gap in the quark spectrum,
along the global minima lines,
is given by $\Delta_3$ in~\eqref{eq:delta_3_squa}.
We show $\Delta_3$  
versus $a/f_a$ in Fig.~\ref{fig:gap3}.
We notice that $\Delta_3$ is a periodic
function of $a/f_a$,
with a period equal to $\pi$,
as expected from the results on $\Delta_L$ and $\Delta_R$ shown 
Fig.~\ref{Fig:gaps}.
We find that although the response of $\Delta_L$ and $\Delta_R$
on $a/f_a$ depends on $\zeta$, hence on the weight of the {$U(1)_A$-breaking}
interaction term, the qualitative behavior of $\Delta_3$ does not
depend on $\zeta$.

\subsection{Axion potential}
 
\begin{figure}[t!]
    \centering
    \includegraphics[scale=0.18]{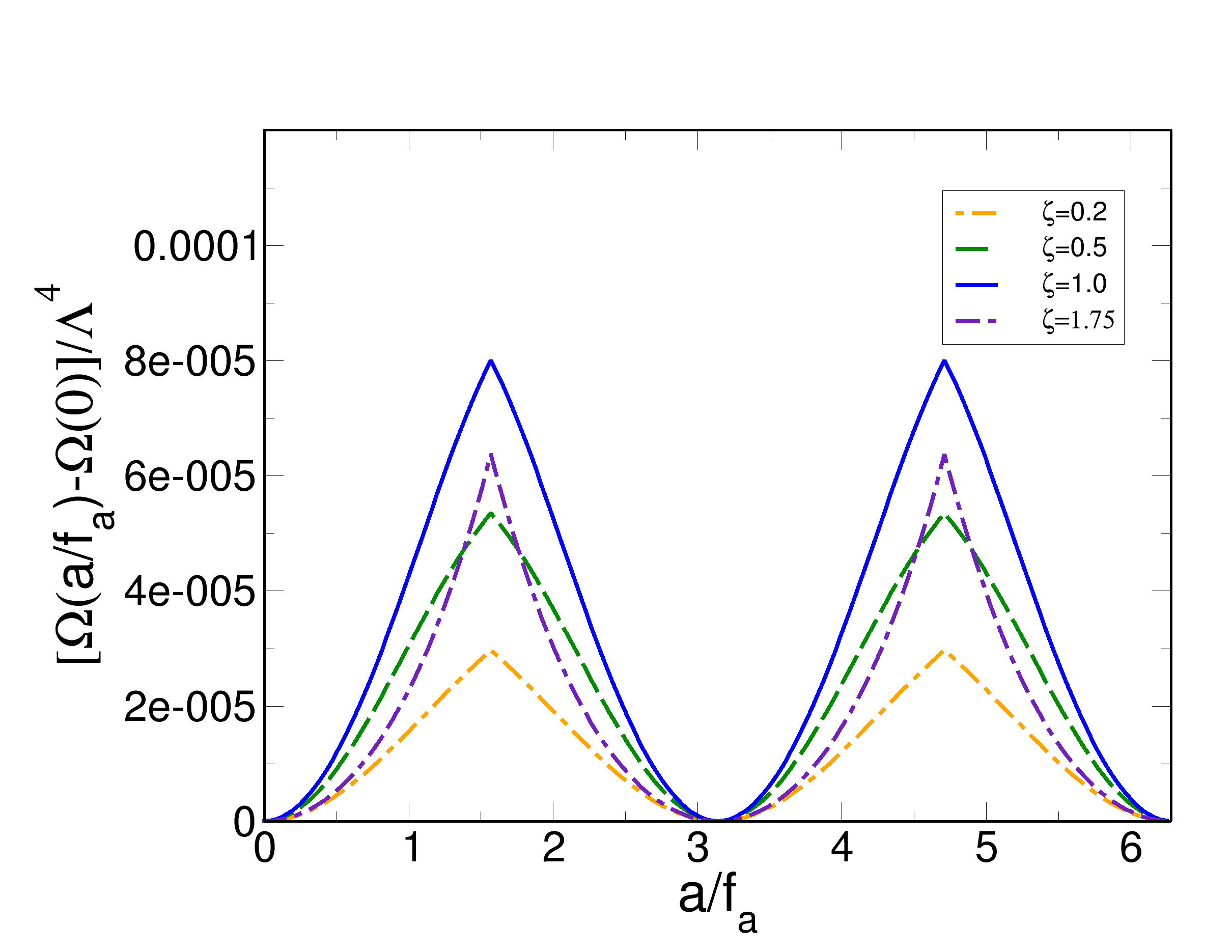}
    
    \caption{Axion potential,
    $V(\theta)=\Omega(\theta)-\Omega(0)$ with
    $\theta=a/f_a$, versus $a/f_a$, computed 
    at 
    $T=0$ and $\mu=400$ MeV
    and for several values of $\zeta$.
    The potential is measured in units of $\Lambda^4$ with $\Lambda=1$ GeV.
    $G_D$ is fixed to have $\Delta_L=50$
    MeV at $a=0$.}
    \label{Fig:omega}
\end{figure}

One of the main results of our work is the computation of the
axion potential, $V(a/f_a)$, in a superconductive phase of QCD.
In Fig.~\ref{Fig:omega} we plot
the $V(a/f_a)\equiv\Omega(a/f_a)-\Omega(a/f_a=0)$ versus $a/f_a$,
at $T=0$ and $\mu=400$ MeV.
For each value of $a/f_a$, the potential
has been computed at the global 
minimum in the $(\Delta_L,\Delta_R)$
space. 
The behavior of $V$ is in agreement with that of the
quark spectrum shown in Fig.~\ref{fig:gap3}, particularly 
for what concerns the periodicity.
We notice that $a=0$ is a global minimum
of $V(a/f_a)$ for the whole range of parameters studied here.
Other degenerate minima are located 
at $a/f_a=n\pi/2$ with $n=\pm1,\pm 2,\dots$.

It is useful to stress the difference with respect
to the axion potential computed in presence
of the chiral and the 
$\eta-$condensate~\cite{Zhang:2023lij} {at zero barion density}:
indeed, in that case, the periodicity of 
$V(a/f_a)$ is equal to $2\pi$, while in the case
of superconductive matter, we find a periodicity
equal to $\pi$. This can be
understood easily. In fact, 
in the present case
at $a=0$ only the scalar condensate forms,
then increasing the value of $a$ the global
minima {become less shallow} and
at $a/f_a=\pi/2$ the thermodynamic potential
develops four degenerate minima, see 
Fig.~\ref{Fig:potential_comparison};
further increasing $a/f_a$ in the range $(\pi/2,3\pi/2)$ results in the 
rotation of the global minima of $\Omega$ 
and in the consequent formation 
of the pseudo-scalar
condensate.
The ground states at $a=0$ and
$a/f_a=\pi$ are degenerate, 
hence the periodicity of the potential,
but the two adjacent minima correspond
to two different condensation channels:
in particular, the ground state at $a=0$ is characterized by
scalar condensation, while the minimum at $\theta=\pi/2$ corresponds
to condensation in the pseudo-scalar channel.

\subsection{Topological susceptibility}

\begin{figure}[t!]
    \centering
\includegraphics[scale=0.18]{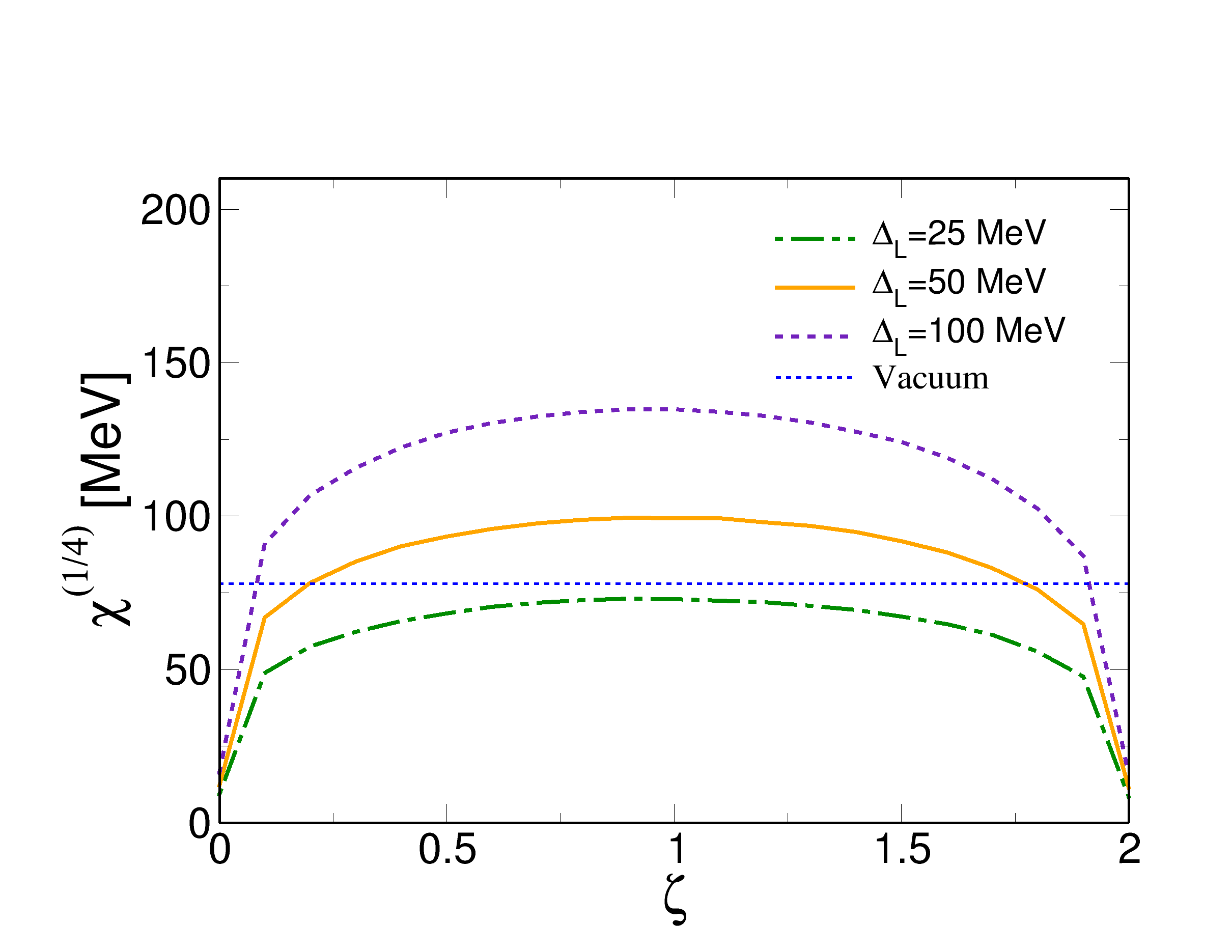}\\
\includegraphics[scale=0.18]{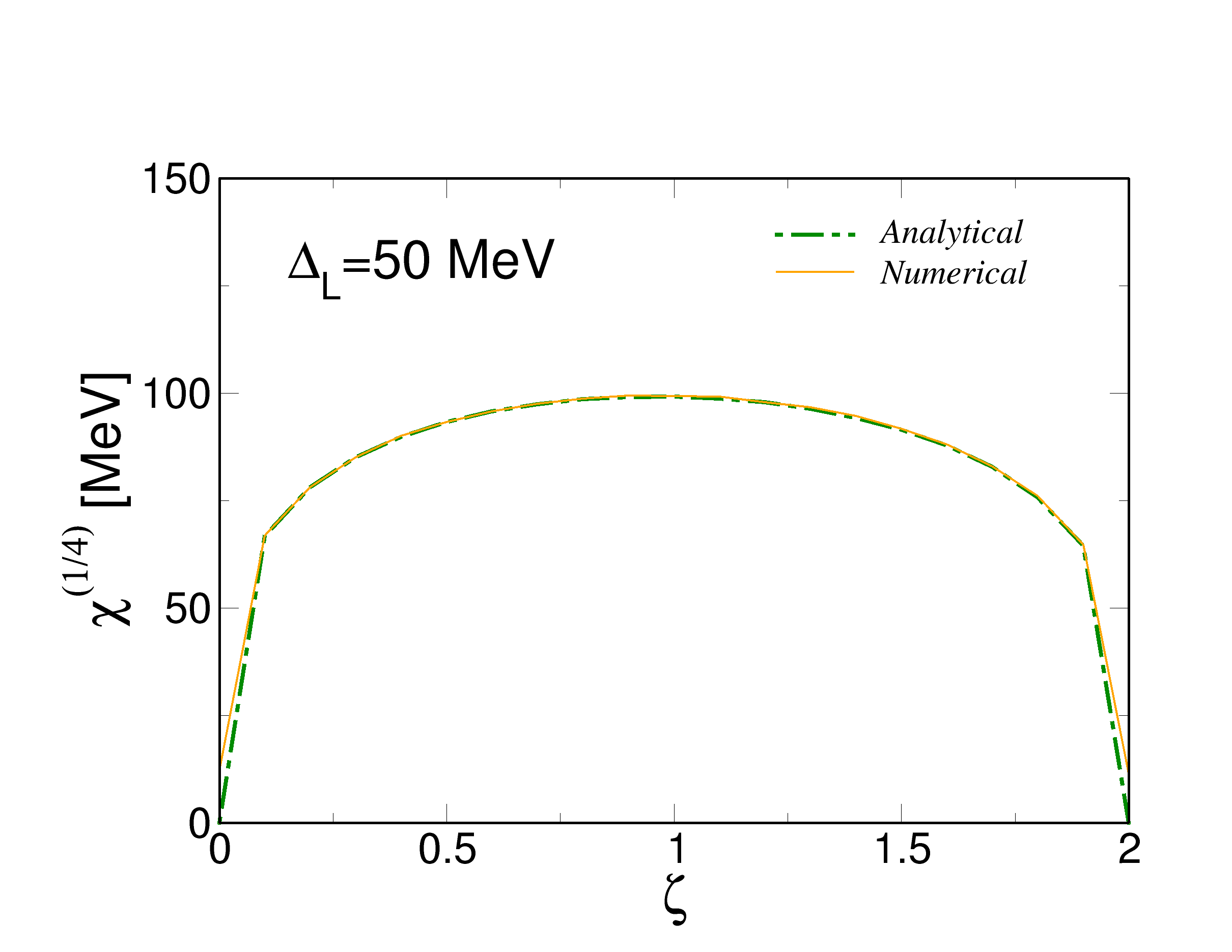}
\caption{Topological
susceptibility, $\chi^{1/4}$, versus $\zeta$, 
 at $T=0$ and
for several values of $\Delta_L$. We used $\mu=400$ MeV,
$\Lambda=1000$ MeV. The value of $\Delta_L$ in the legend has been used
to fix the value of $G_D$, so that for each line $\Delta_L$
is kept fixed while $\zeta$ is changed. {The horizontal blue line corresponds to the reference value expected for the topological susceptibility in the vacuum, see, e.g.,\cite{Lu:2018ukl,Borsanyi:2016ksw}. }
The analytical curve corresponds to Eq.~\eqref{eq:hacommessolerroredinondirtelo}
computed for 
$\Delta_L=50$ MeV and $G_D\Lambda^2=4.69$.}
    \label{fig:eracroce}
\end{figure}

In this section, we analyze the 
full topological
susceptibility,  $\chi$, which measures the fluctuations
of the topological charge and is defined as
\begin{equation}
\chi=\left.\frac{d^2\Omega}{d\theta^2}\right|_{\theta=0},
~~~\theta=\frac{a}{f_a}.
\label{eq:vedetemicapicacontinuamente}
\end{equation}
In the numerical calculation, we treat the above derivative
as a total derivative, that in principle takes into account
also of the dependence of the condensates on $\theta$
at $\theta=0$.

In Fig.~\ref{fig:eracroce} 
we plot the fourth root of $\chi$
versus $\zeta$ , 
 at $T=0$ and
for several values of $\Delta_L$. We used $\mu=400$ MeV,
$\Lambda=1000$ MeV. The value of $\Delta_L$ in the legend has been used
to fix the value of $G_D$, so that for each line $\Delta_L$
is kept fixed while $\zeta$ is changed.
$\chi$ is an interesting quantity by itself,
since it encodes information about the fluctuations of
the topological charge in the dense and superconductive
QCD medium. Moreover, it is directly related to the
squared axion mass, see the next subsection.
Previous studies performed within Lattice QCD in the
isospin-symmetric case, 
as well as within 
chiral perturbation theory and NJL models, 
agree on the value $\chi^{1/4}\approx 78$ MeV
at $T=0$ and $\mu=0$~\cite{Borsanyi:2016ksw,GrillidiCortona:2015jxo,Lu:2018ukl,Gatto:2011wc},
as well as with the Di Vecchia-Crewther-Leutwyler-Smilga-Veneziano formula~\cite{Veneziano:1979ec,DiVecchia:1980yfw,Crewther:1977ce,Leutwyler:1992yt},
that for two flavors reads
\begin{equation}
\chi=|\langle\bar q q\rangle|\frac{m_u m_d}
{m_u+m_d},\label{eq:DVC}
\end{equation}
where $\langle\bar q q\rangle$ is the chiral 
condensate in the vacuum.
$\chi$ then decreases at high $T$ where the smooth
crossover to the quark-gluon plasma phase takes place~\cite{Borsanyi:2016ksw,GrillidiCortona:2015jxo,Lu:2018ukl,Gatto:2011wc}.

The results in Fig.~\ref{fig:eracroce}
show that for a given value of $\zeta$,
the topological susceptibility increases with the
strength of the coupling, as expected.
Moreover, if the coupling is tuned in order to
give a value of $\Delta_L$, changing $\zeta$
within the range $(0.5,1.5)$ 
does not substantially affect $\chi^{1/4}$.
We also note that keeping $\zeta$ in the aforementioned
range keeps $\chi^{1/4}$ in the superconductive phase
in the same ballpark of the value it takes in the
vacuum, unless we take
a very large superconductive gap 
as in the case $\Delta_L=100$ MeV shown in 
Fig.~\ref{fig:eracroce}.

Within our model
we were able to obtain an analytical result for $\chi$.
In fact,
taking into account that at the global minimum
$\partial\Omega/\partial\Delta_L=\partial\Omega/\partial
\Delta_R=0$, and according to the results in
Fig.~\ref{Fig:gaps} we have 
$\partial\Delta_L/\partial \theta=
\partial\Delta_R/\partial \theta=0$ at $\theta=0$
(see also Eq.~\eqref{eq:gentechegridasulbus}),
it is straightforward to prove that (see appendix~\ref{sec:equationDERIVATION})
\begin{equation}
\chi = \frac{\Delta_L^2}{2G_D}\zeta
\frac{2-\zeta}{2+\zeta},
\label{eq:hacommessolerroredinondirtelo}
\end{equation}
where  
$\Delta_L$  
corresponds to the solution of the gap equation at $\theta=0$; 
Eq.~\eqref{eq:hacommessolerroredinondirtelo} stands 
both at zero and at finite temperature.
{In the bottom panel of Fig. 6 we plot the fourth root of the topological
susceptibility versus $\zeta$, obtained by Eq.~\eqref{eq:hacommessolerroredinondirtelo} using $\Delta_L$= 50 MeV and $G_D\Lambda^2 =$ 4.69.
The agreement between the analytical result and the numerical one is
self-explanatory (we checked the agreement also for other values of the
parameters).}

\subsection{Axion mass and self-coupling}

The low energy lagrangian density of the 
axion field can be written as
\begin{equation}
\mathcal{L}_a = \frac{1}{2}\partial^\mu \partial_\mu a -\frac{m_a^2}{2} a^2 -\frac{\lambda_a}{4!} a^4,
\label{eq:lalagrangiana_assione}
\end{equation}
where the axion mass and the quartic coupling
are defined in terms of the potential $V(\theta)=\Omega(\theta)-\Omega(0)$ as
\begin{equation}
m_a^2=
\frac{1}{f_a^2}
\left.\frac{d^2\Omega}{d\theta^2}\right|_{\theta=0},~~~
\lambda_a  = \frac{1}{f_a^4}
\left.\frac{d^4 V(\theta)}{d\theta^4}\right|_{\theta=0},
\label{eq:noveperdieciallasei}
\end{equation}
and $\theta=a/f_a$.
Within our model, we can compute how
the axion mass and coupling behave in the 
superconductive phase of QCD, as well as 
study their response to the phase transition to
the normal phase.

From~\eqref{eq:laportasiapre}
we get an analytical formula for the axion-squared mass
in the color-superconductive phase, which is
\begin{equation}
m_a^2  = \frac{\Delta_L^2}{2G_D f_a^2}
|\zeta|\frac{2-|\zeta|}{2+|\zeta|},
\label{eq:nelcuoredelledolomiti}
\end{equation}
where $\Delta_L$ corresponds to the solution
of the gap equation.
To our knowledge, Eq.~\eqref{eq:nelcuoredelledolomiti} is
a new result in the literature.

\begin{figure}[t!]
    \centering
    \hspace{-13mm}\includegraphics[scale=0.18]{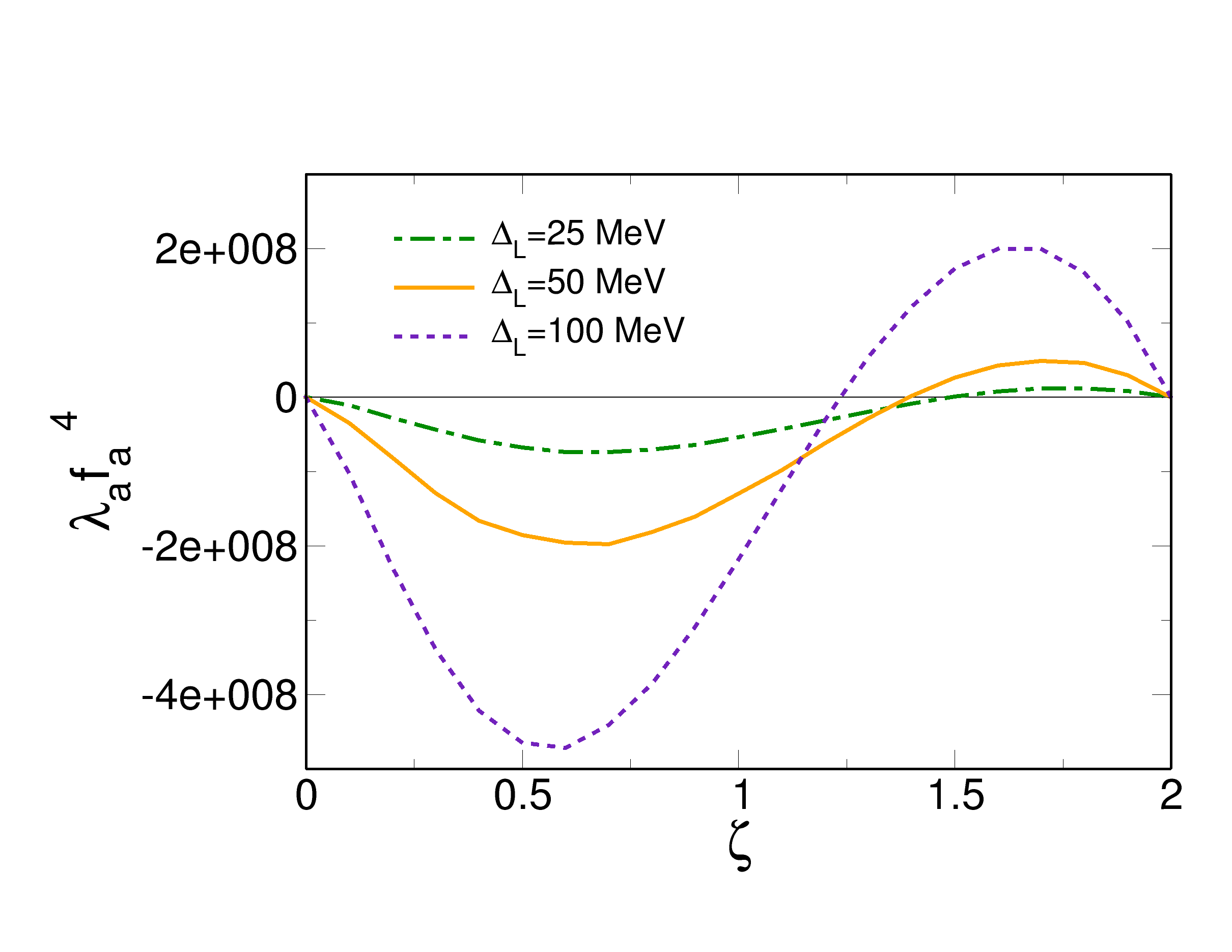}
\caption{Axion self-coupling, $\lambda_a f_a^4$, versus  $\zeta$ for several values of $\Delta_L$.}
    \label{fig:lam}
\end{figure}

We have not been able to find an analytical expression for
$\lambda_a$, hence we computed it numerically.
In Fig.~\ref{fig:lam} we plot the rescaled
axion self-coupling,
versus $\zeta$, for several values of $\Delta_L$.
The parameters are the same as used in Fig.~\ref{fig:eracroce}.
For the sake of comparison,
we note that studies based on the NJL model
find $\lambda_a f_a^4=-(55.64$ MeV$)^4$
at $T=\mu=0$~\cite{Lu:2018ukl}.
We find that
$\lambda_a$ is negative in a range of 
$\zeta$ that partly depends on $\Delta_L$,
hence on the strength of the coupling.
This sign of $\lambda_a$ is in agreement 
with what was found within NJL models
at zero as well as finite $\mu$~\cite{Lu:2018ukl,Zhang:2023lij}.
However, we find also a range of
$\zeta$ where $\lambda_a$ is positive.
The value $\bar\zeta$ 
of $\zeta\neq 0$ such that
$\lambda_a=0$ depends on the strength
of the coupling: however, comparing
with the results shown in Fig.~\ref{Fig:gaps},
we find that 
$\bar\zeta$ is in agreement with the
value at which $\Delta_L$ and $\Delta_R$ 
invert their tendency to change as
$a/f_a$ is increased, see Eq.~\eqref{eq:sembrachesiaunpensierocomune}.

\subsection{Finite temperature}

\begin{figure}
    %\centering
    \hspace{-15mm}\includegraphics[scale=0.18]{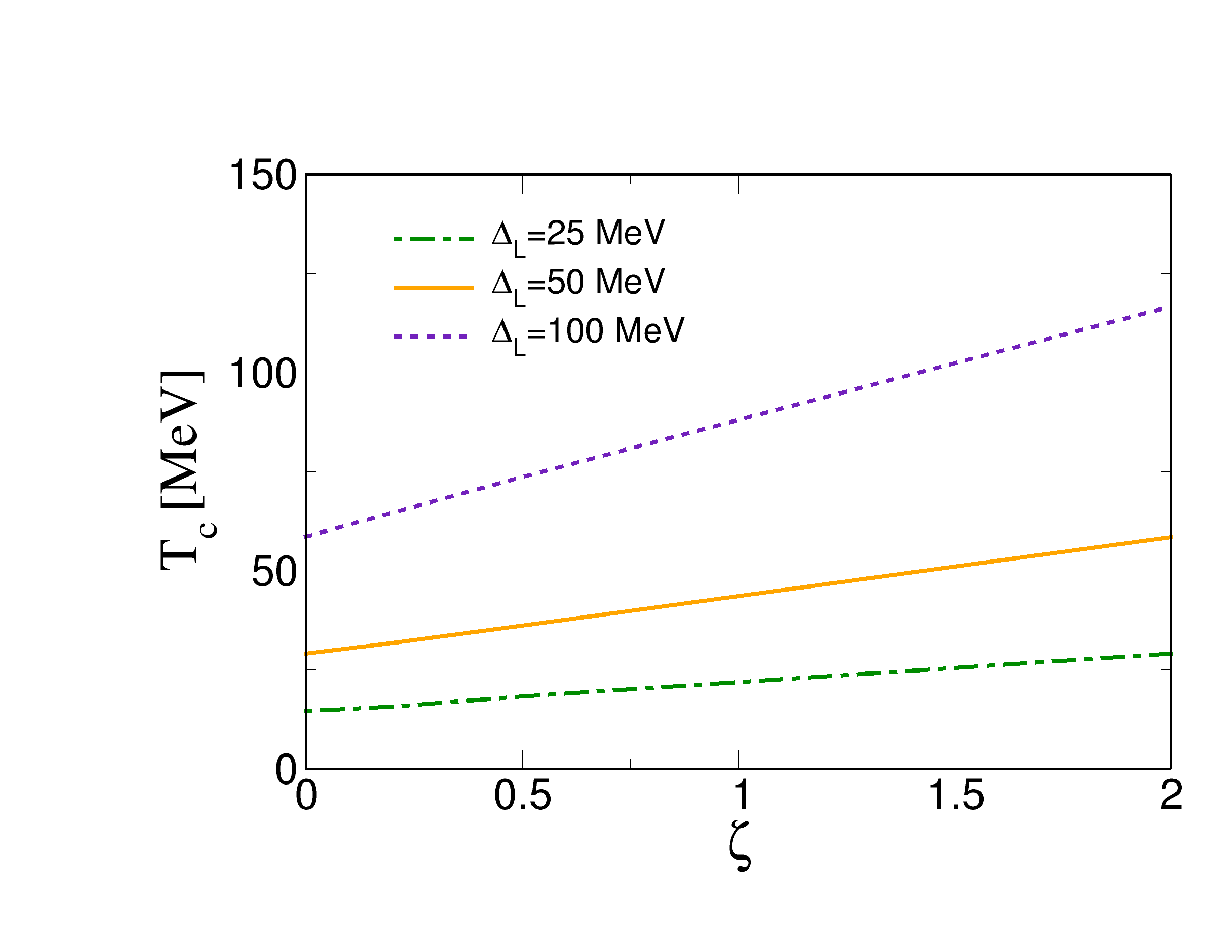}
    \caption{Critical temperature, $T_c$, as a function of $\zeta$, for different values of $\Delta_L$. Calculations correspond to $a=0$.}
    \label{Fig:temp_crit_fm}
\end{figure}

In this section, we 
briefly investigate the axion potential in dense superconductive
quark matter at a small, albeit finite, temperature.
Increasing the temperature we expect a critical value, depending on the chemical potential, for which a phase transition occurs from the color-superconducting phase to normal-quark matter. In particular, when the critical temperature $T_c$ is reached, the gaps $\Delta_L$ and $\Delta_R$ vanish and a second-order phase transition occurs.
We do not push this investigation too much, because
we neglected quark masses in this work, that contribute
to obtain nonzero values of  $\chi$ and $\lambda_a$ above the
critical temperature: we limit ourselves to compute how $T_c$
depends on $\zeta$, as well as to show the behavior of $\chi$
and $\lambda_a$ around $T_c$ obtained within our model.

In Fig.~\ref{Fig:temp_crit_fm} we plot
$T_c$ 
versus $\zeta$ for several values of $\Delta_L$; 
$G_D$ is varied {as a function of} $\zeta$, such that along each line the value of $\Delta_L$ at $T=0$ is kept constant. 
We find that $T_c$ increases linearly as a function of $\zeta$;
moreover,
$T_c$ increases upon increasing $\Delta_L$ as expected. This is in agreement with the analytic result that can be found 
within the framework of the HDET, 
\begin{equation}
    T_c=\Delta_L^0\frac{e^\gamma (2+\zeta)}{2\pi}\,,
   \label{deltatc}
\end{equation}
where $\Delta_L^0$ is the value of the gap at vanishing temperature and $\gamma\approx0.577$ 
is the Euler-Mascheroni constant. See Appendix~\ref{sec:deltatcDERIVATION}
for a derivation of~\eqref{deltatc}.

\begin{figure}[t!]
    \centering
    \includegraphics[scale=0.18]{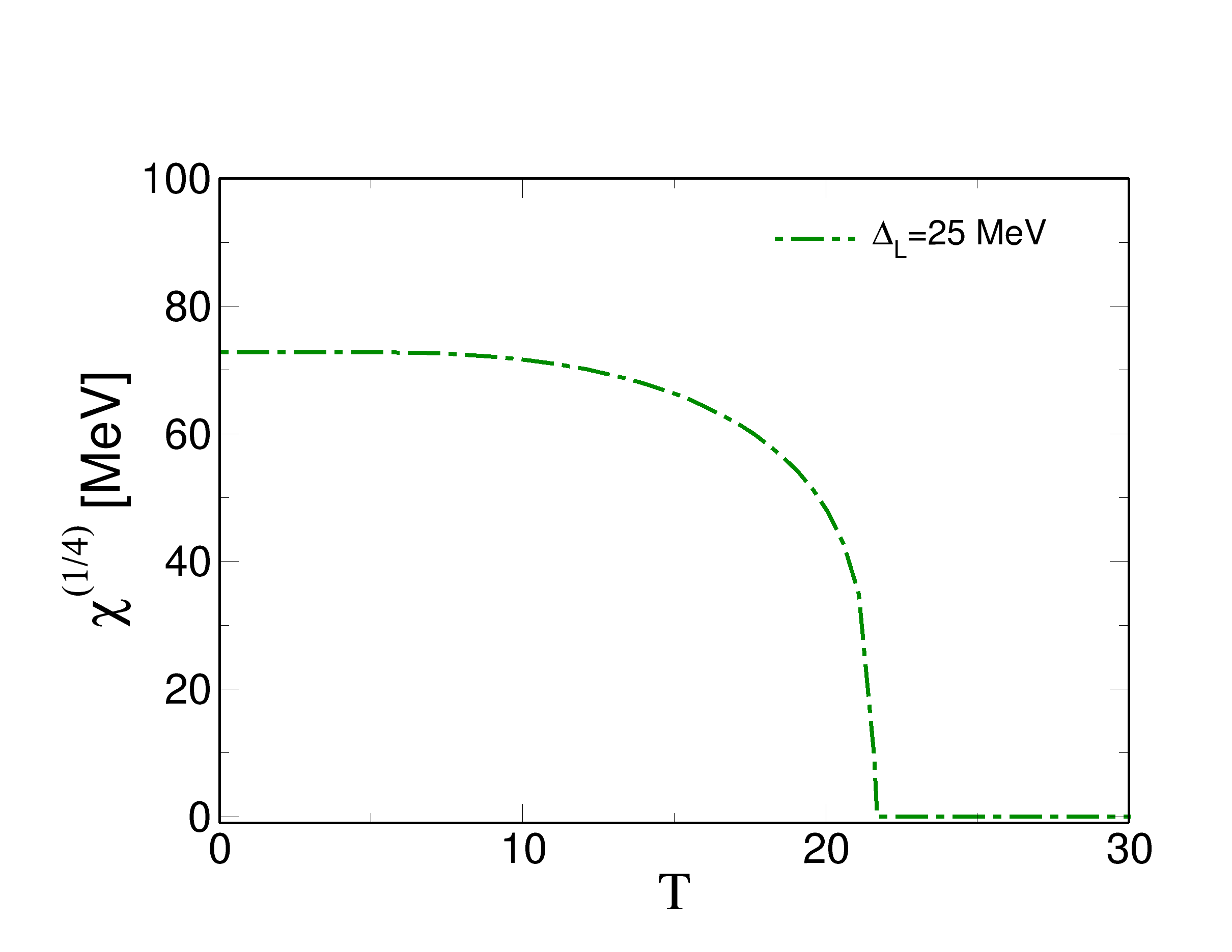}\\
    \hspace{-7mm}\includegraphics[scale=0.18]{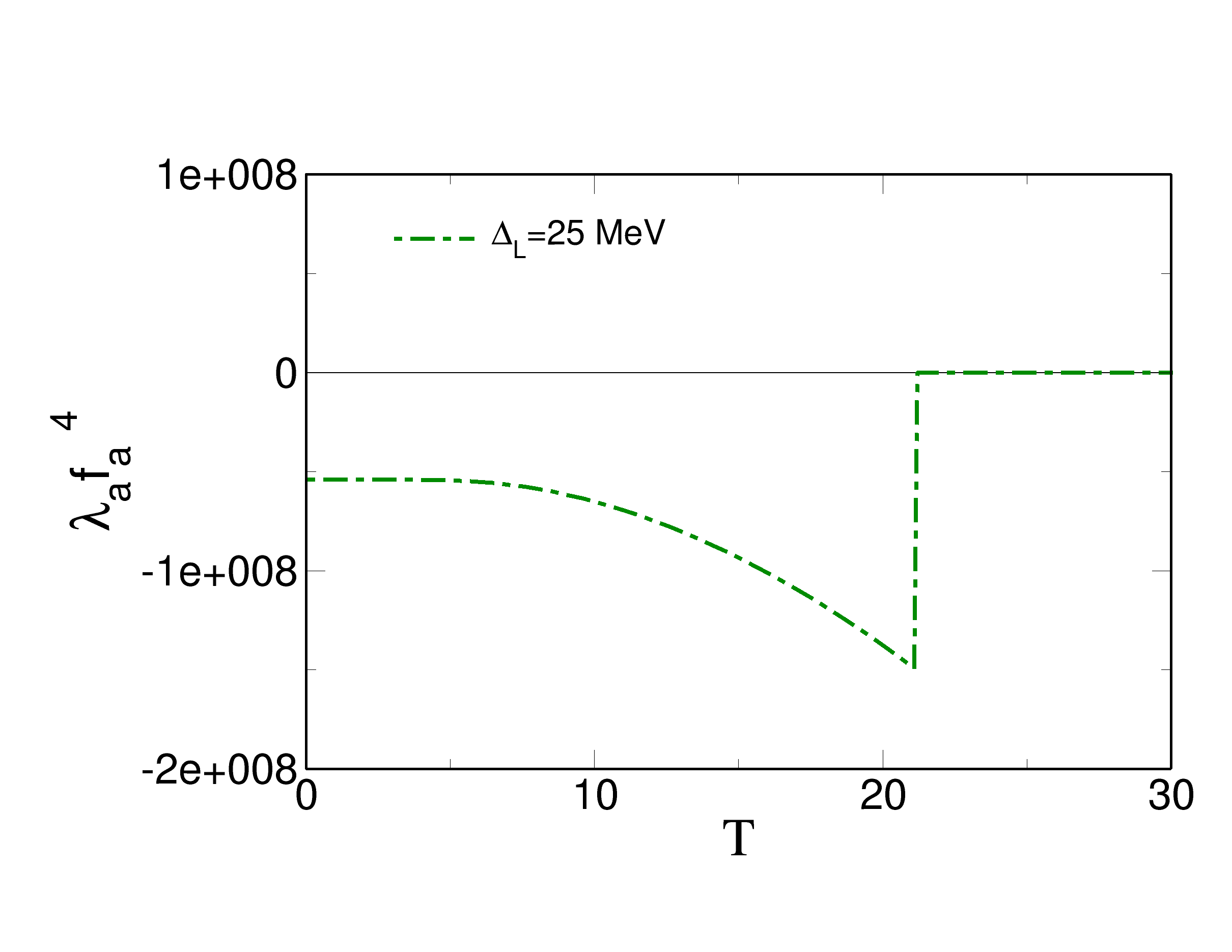}
\caption{Topological
susceptibility, $\chi^{1/4}$ (upper panel) and axion self-coupling, $\lambda_a f_a^4$ (lower panel), versus  $T$ for $\zeta=1$. We used $\mu=400$ MeV,
$\Lambda=1000$ MeV and  $\Delta_L=25$ MeV.}
    \label{fig:lamT}
\end{figure}

It is also interesting to check the behavior of
$\chi^{1/4}$ and $\lambda_a$ versus $T$
near the transition to normal quark matter.
In particular, we verified that \cref{eq:hacommessolerroredinondirtelo} is also valid in the finite-temperature case
(see appendix~\ref{sec:equationDERIVATION}), and the temperature dependence enters only via $\Delta_L$. 
In the upper panel of \cref{fig:lamT} we show the topological
susceptibility, $\chi^{1/4}$ (upper panel) and axion self-coupling, $\lambda_a f_a^4$ (lower panel), versus  $T$ for $\zeta=1$. We used $\mu=400$ MeV and  $\Delta_L=25$ MeV. 
The trend we find is in agreement with
\cref{eq:hacommessolerroredinondirtelo},
with a vanishing $\chi$ and $\lambda_a$ in the normal phase.
However, we remark that the vanishing of these
two quantities in the normal phase is an artifact of our
one-loop approximation to the thermodynamic potential,
as well as neglecting the quark masses:
including quark masses would lead to nonzero
axion mass and self-coupling also in the normal phase,
similarly to what happens in the high temperature/low density
region of the QCD phase diagram~\cite{Lu:2018ukl,Zhang:2023lij,GrillidiCortona:2015jxo}.

\section{Conclusions and outlook}
We analyzed
the QCD axion potential in dense,
superconductive quark matter
at finite quark chemical potential, $\mu$,
and finite temperature, $T$.
We assumed that both scalar, $\Delta_R-\Delta_L$, 
and pseudoscalar, $\Delta_R + \Delta_L$,
diquark condensates can form: the color-flavor
structure that we assume is that of the standard
2SC phase.
We used a two-flavor model-based 
on an interaction that contains a $U(1)_A$-preserving
and a $U(1)_A$-breaking term: the former 
has coupling strength $G_D$ and 
is assumed
to be derived from one-gluon exchange, while the
latter arises from the one-instanton exchange,
it has coupling strength $\zeta G_D$, and
contains the coupling of the QCD axion, $a$, to the
quarks. We treated $G_D$ and $\zeta$ as free
parameters: in particular, we tuned $G_D$ for
a given $\zeta$ in order to reproduce a value
of the superconductive gap without axions;
we then switch on the axion field and study the
response of the diquark condensate, as well as
of the thermodynamic potential, to $a$.
This allowed us to compute the axion potential
in dense quark matter.

We found that for $a=0$ and $\zeta> 0$ the
scalar condensate is favored; however,
increasing $a/f_a$
in the range $(0,\pi/2)$ results 
in the the global minima 
of $\Omega$ {to become less shallow} along the direction of the scalar 
condensate and in the formation of new global
minima around the direction of the pseudoscalar. Consequently, for 
$a/f_a$
in the range $(\pi/2, 3\pi/2)$ the pseudoscalar
condensate replaces the scalar one.
Increasing further $a/f_a$ up to $2\pi$ 
the location of the global minima of $\Omega$
change again and the system turns to condensate
in the scalar channel.

We computed the full topological susceptibility, $\chi$,
of the 2SC phase, finding an analytical formula that
connects $\chi$ to the quark condensate; this
relation holds both at zero and at finite temperature.
Our formula~\eqref{eq:hacommessolerroredinondirtelo} explicitly contains
information relative to the specific model used 
in our work, in particular, the effective coupling
$G_D$ and $\zeta$. It is likely that in a more
refined framework, in which one uses a
momentum-dependent
one-gluon exchange term and an
instanton kernel instead of our effective
$4-$fermion interactions, both $G_D$ and $\zeta$
will be replaced by quantities directly
related to QCD, namely the QCD coupling,
a dressed gluon mass as well as the instanton
size. This interesting improvement will be the
the subject of a near future work. {The introduction of the momentum dependent gluon term would  also allow us to explore the effect of cutting high momenta in the temperature dependent contribution to the thermodynamic potential,  as we mentioned in Section \ref{sec:ld}.}

We then computed
the axion mass, $m_a$,
which is related to $\chi$ by the
relation $\chi=m_a^2 f_a^2$:
as a consequence of our
result~\eqref{eq:hacommessolerroredinondirtelo}, we were able to
obtain an analytical formula for $m_a$
in the 2SC phase of QCD. Also, in this case, it is our
hope that the dependence of $m_a$ on the parameters
of our model can be replaced by a dependence on 
quantities of QCD by using more refined interactions
instead of the $4-$fermion terms we adopted here.
We completed this part of the study by computing
the axion self-coupling, $\lambda_a$. 
Interestingly, we found a range of $\zeta$ where
$\lambda_a>0$, differently from what it was found
in cases of normal quark matter, see for example~\cite{Lopes:2022efy,Bandyopadhyay:2019pml,Das:2020pjg,Lu:2018ukl,Zhang:2023lij,GrillidiCortona:2015jxo}:
hence the interaction among axions becomes repulsive
in this range of $\zeta$. This might have some impact
on the pressure of axions trapped in the core
of compact stellar objects. It will be interesting 
to explore this scenario in detail in the future.

A possible improvement 
of the present work
is 
the inclusion
of the strange quark, and the opening to the
three-flavor color superconductor with massive quarks.
This would potentially make the picture
more complicated, because in this work
we neglected the current quark masses,
as well as light quarks 
chiral condensate: previous studies
based on NJL-like interactions show that
this can be a fairly good approximation, as long as
the light-quarks sector is considered. 
However, this might be no longer true for the 
strange quark, at least for the values of $\mu$ which
are relevant for compact stars.
Along this line, studying the axions in gapless phases
is also worth more investigation: works in this direction
are already ongoing and we plan to report on them soon.
Secondly, it would be interesting to 
use the results obtained here to compute
the cooling of compact stellar objects via
axion emission.
It would also be important to check how the
picture we drew in our work changes when 
the local interaction kernels we used are replaced
by non-local ones, with a more direct link
to QCD, as well as when
other interaction channels, in primis the vector
and axial-vector channels are included in the game.
Even more, it is of a certain interest to allow for
chiral condensate besides the diquark one, 
in order to study, 
within a single model, the axion potential
in proximity of the phase transition
between the chiral and the superconductive phases:
this would allow for the exploration of the properties
of the axion in the whole QCD phase diagram.
We leave all these interesting 
improvements
to future works.

\begin{acknowledgements}
M. R. acknowledges John Petrucci for inspiration.
D.E.A.C. acknowledges support from the program Excellence Initiative–Research University of the University of Wroclaw of the Ministry of Education and Science. A.G. G. would like to acknowledge the support received from CONICET (Argentina) under Grants No. PIP 22-24 11220210100150CO and from ANPCyT (Argentina) under Grant No. PICT20-01847.
This work has been partly funded by the
European Union – Next Generation EU through the
research grant number P2022Z4P4B “SOPHYA - Sustainable Optimised PHYsics Algorithms: fundamental
physics to build an advanced society” under the program
PRIN 2022 PNRR of the Italian Ministero dell’Università
e Ricerca (MUR).

\end{acknowledgements}

\appendix
\section{HDET gap equation\label{sec:appendix:hdet}}
In this section, we derive the HDET solution
to the
gap equation~\eqref{eq:matchingorganicmatter}.
To this end, we put
\begin{equation}
\Theta(a,\zeta) = \frac{A(a,\zeta)}
{2 + \zeta \cos(a/f_a)}.
\label{eq:ameteoritefrommars}
\end{equation}
Then, Eq.~\eqref{eq:matchingorganicmatter} gives
\begin{equation}
1 = 
\frac{G_D}{2\pi^2}\Theta(a,\zeta)\int_0^\Lambda p^2dp~
\left(
\frac{1}{\sqrt{(p-\mu)^2 + G_D^2 h_L^2 A(a,\zeta)}}
+
\frac{1}{\sqrt{(p+\mu)^2 + G_D^2 h_L^2 A(a,\zeta)}}
\right).
\label{eq:tidalforcesoneuropa}
\end{equation}
We then 
adopt the approximations of HDET~\cite{Nardulli:2002ma}. Firstly, we note that the 
the first integral on the right-hand side
of~\eqref{eq:tidalforcesoneuropa} gets its largest contribution
from the momentum space region $p\approx\mu$,
namely around the Fermi surface of the quarks.
Moreover, the second integral in the right-hand side
is suppressed at large $\mu$ in comparison with the first one,
since it does not receive the enhancement for $p\approx\mu$.
Within the spirit of HDET we can thus ignore the second
integral, and restrict the first integral to a 
thin shell around
$p=\mu$: we call $\delta$ the width of this shell.
In the limit $\delta\ll\mu$ the volume of momentum space
available for pairing is thus $8\pi\mu^2\delta$.
Introducing $\xi=p-\mu$,
the HDET version of~\eqref{eq:tidalforcesoneuropa} reads\footnote{To be precise,
in the HDET one introduces a sum over the direction of the Fermi velocities, $\bm v_F$, 
of the
quarks; then, for each $\bm v_F$,
$\xi=\bm p \cdot \bm v_F - \mu v_F$ 
measures the fluctuation of the longitudinal momentum around the
Fermi surface. Our $\xi$ in~\eqref{eq:troppamusicabanale} is slightly different
from that of HDET, nevertheless, formally the gap equation obtained within HDET
is in agreement with~\eqref{eq:troppamusicabanale},
because the quark condensate is homogeneous and the sum over
velocities lead to an overall $1$.}
\begin{equation}
1 = 
\frac{G_D \mu^2}{2\pi^2}\Theta(a,\zeta)\int_{-\delta}^{+\delta}
\frac{d\xi}{\sqrt{\xi^2 + G_D^2 h_L^2 A(a,\zeta)}}.
\label{eq:troppamusicabanale}
\end{equation}
Integration can be done exactly;
in the weak coupling
limit $\delta\gg G_D h_L$, 
using the definition~\eqref{eq:def_capital_DDD_aa},
we finally get
\begin{equation}
\Delta_L = \frac{4\delta}{\sqrt{A(a,\zeta)}}
\exp\left(
-\frac{\pi^2}{\mu^2 G_D \Theta(a,\zeta)}
\right).
\label{eq:sparicomeuna}
\end{equation}

\section{Derivation of Eq.~\eqref{eq:hacommessolerroredinondirtelo}\label{sec:equationDERIVATION}}
In this section, we derive Eq.~\eqref{eq:hacommessolerroredinondirtelo}
starting from $\Omega$ in~\eqref{eq:Omega_kkk_ll_ppp}
and using the gap equation~\eqref{eq:matchingorganicmatter}.
The first step in the calculation is to notice that 
all the results we found are consistent with the conditions
$\partial\Omega/\Delta_L = \partial\Omega/\Delta_R=0$ at $\theta=0$
and at the minima of $\Omega$. This can be easily understood since
$\Omega$ develops minima along the directions $\Delta_L = \pm\Delta_R$,
and on these lines, it is an even function of $\Delta_L$ or $\Delta_R$.
Hence we can write
\begin{equation}
\chi=
\left.\frac{d^2\Omega}{d\theta^2}\right|_{\theta=0}
=\left.\frac{\partial^2\Omega}{\partial\theta^2}\right|_{\theta=0}.
\label{eq:stazittovecchioscemo}
\end{equation}
From Eq.~\eqref{eq:Omega_kkk_ll_ppp} we get
\begin{equation}
\chi=2\zeta G_D h_L h_R -\frac{1}{\pi^2}\int_0^\Lambda p^2dp
\frac{\partial^2}{\partial\theta^2}\left(
\varepsilon_3 +\varepsilon_5 + \mu\rightarrow-\mu
\right).
\label{eq:musicastranapiugwen}
\end{equation}
Along the line $h_L = -h_R$, which is the relevant one for small $\theta$, and taking into account the expressions of the dispersion
laws of the quarks,
we thus get
\begin{equation}
\chi=8\zeta G_D h_L^2\left[-\frac{1}{4} +
\frac{G_D}{2\pi^2}\int_0^\Lambda p^2dp
\left(
\frac{1}{\sqrt{(p-\mu)^2 + (2+\zeta)^2G_D^2 h_L^2}}
+ \mu\rightarrow-\mu
\right)\right].
\label{eq:sologridaepianti}
\end{equation}
Now, we notice from the gap equation~\eqref{eq:matchingorganicmatter} at
$\theta=0$ that
\begin{equation}
\frac{1}{2+\zeta} = 
\frac{G_D}{2\pi^2}\int_0^\Lambda p^2dp
\left(
\frac{1}{\sqrt{(p-\mu)^2 + (2+\zeta)^2G_D^2 h_L^2}}
+ \mu\rightarrow-\mu
\right),
\label{eq:glizombiesonoorribili}
\end{equation}
where $h_L$ stands for the condensate at $\theta=0$.
Using~\eqref{eq:glizombiesonoorribili} in~\eqref{eq:sologridaepianti}
we get
\begin{equation}
\chi=2 G_D h_L^2 
\zeta \frac{2-\zeta}{2+\zeta}.
\label{eq:noncisonocartucc}
\end{equation}
Finally, taking into account $\Delta_L = 2G_D h_L$
(see Eq.~\eqref{eq:def_capital_DDD_aa}) we have
\begin{equation}
\chi=\frac{\Delta_L^2}{2G_D} \zeta
\frac{ 2-\zeta }{2+\zeta},
\label{eq:ashumanityevolved}
\end{equation}
in agreement with
Eq.~\eqref{eq:hacommessolerroredinondirtelo}.
We remark that $\Delta_L$ in~\eqref{eq:ashumanityevolved}
denotes the diquark condensate at $\theta=0$.
From the derivation presented here, it is evident that
Eq.~\eqref{eq:hacommessolerroredinondirtelo} stands both at
zero and at finite temperature, since the momentum integrals
in Eqs.~\eqref{eq:sologridaepianti}
and~\eqref{eq:glizombiesonoorribili} are modified in the
same fashion at $T\neq 0$.

As we remarked in the main text, 
for $\zeta<0$ the role of the
scalar and pseudoscalar condensates invert, since the minima of $\Omega$
develop along the line $h_L=h_R$ in this case,
and Eq.~\eqref{eq:hacommessolerroredinondirtelo} becomes
\begin{equation}
\chi = -\frac{\Delta_L^2}{2G_D}\zeta
\frac{2+\zeta}{2-\zeta}.
\label{eq:hacommessolerroredinondirtelo_negzet}
\end{equation}
Hence, we can summarize the results~\eqref{eq:hacommessolerroredinondirtelo}
and~\eqref{eq:hacommessolerroredinondirtelo_negzet} as
\begin{equation}
\chi = \frac{\Delta_L^2}{2G_D}|\zeta|
\frac{2-|\zeta|}{2+|\zeta|}.
\label{eq:laportasiapre}
\end{equation}

\section{Derivation of Eq.~\eqref{deltatc}\label{sec:deltatcDERIVATION}} 
We consider the HDET gap equation \cref{eq:troppamusicabanale}, and evaluate the integral analytically in the case $a=0$ and $\delta\gg G_D h_L$ thus obtaining
\begin{equation}
1 = 
\frac{G_D \mu^2}{\pi^2}(2+\zeta)\ln\left(\frac{4\delta}{G_D h_L^0 (2+\zeta)}\right),
\label{gappabella}
\end{equation}
where $h_L^0$ is the value of the gap which satisfies the 0-temperature gap equation. 

In the same framework, we now consider the finite temperature gap equation for $a=0$, which can be written as 
\begin{equation}
-1 + 
\frac{G_D \mu^2}{2\pi^2}(2+\zeta)\int_{-\delta}^{+\delta}
\frac{d\xi}{\sqrt{\xi^2 + G_D^2 h_L^2 (2+\zeta)^2}}=\frac{G_D \mu^2}{2\pi^2}(2+\zeta)\int_{-\delta}^{+\delta}
\frac{d\xi}{\sqrt{\xi^2 + G_D^2 h_L^2 (2+\zeta)^2}\left(1+e^{\sqrt{G_D h_L(2+\zeta)^2+\xi^2}/T}\right)}.
\label{gappatot}
\end{equation}
The integral appearing on the left-hand side
of the equation above is analogous to the one in \cref{eq:troppamusicabanale}. Thus, the integration leads to the same result that appears in the r.h.s of \cref{gappabella}, with the only difference that $h_L^0$ is replaced by the finite temperature gap $h_L$.
Inserting then \cref{gappabella} into the l.h.s. of \cref{gappatot}, 
and noticing that
the fast convergence of the integral
in the right-hand side of~\eqref{gappatot} 
allows us to extend the integration to the whole real axis, we obtain
\begin{equation}
   \ln{\left(\frac{h_L^0}{h_L}\right)}=
   \int_{-\infty}^{+\infty}
\frac{d\xi}{\sqrt{\xi^2 + G_D^2 h_L^2 (2+\zeta)^2}\left(1+e^{\sqrt{G_D h_L(2+\zeta)^2+\xi^2}/T}\right)}\equiv I(u), 
\label{gappa3}
\end{equation}
where we put
\begin{equation}
   I(u)= \int_{-\infty}^{+\infty}
\frac{d x}{\sqrt{x^2 + u^2}\left(1+e^{\sqrt{u^2+x^2}}\right)},
\label{iu}
\end{equation}
with $x=\xi/T$ and $u=G_Dh_L(2+\zeta)/T$.
For $T\rightarrow T_c$ the condensate $h_l\to0$: 
we can thus limit ourselves to 
consider the leading order expansion of \cref{iu} for $u\sim0$, namely
\begin{equation}
I(u)\Big|_{u\to0}\simeq \ln\left(\frac{\pi}{e^\gamma u}\right),
\end{equation}
where $\gamma\approx0.577$ is the Euler-Mascheroni 
constant. Using this result in \cref{gappa3} we obtain
\begin{equation}
\ln{\left(\frac{h_L^0}{h_L}\right)}=\ln{\left(\frac{\pi T_c}{\gamma G_D (2+\zeta) h_L}\right)}, 
\label{gappafin}
\end{equation}
where we set $T=T_c$ and discarded all the finite $h_L^2$ contributions.  
Finally, \cref{gappafin} can be fulfilled only if
\begin{equation}
h_L^0=\frac{\pi T_c}{\gamma G_D (2+\zeta)}, 
\end{equation}
which gives
from which we trivially derive 
\begin{equation}
T_c=\Delta_L^0\frac{e^\gamma (2+\zeta)}{2\pi},
\end{equation}
in agreement with~\eqref{deltatc}.

\end{document}